\documentclass[12pt,preprint]{aastex}
\usepackage{graphicx}
\usepackage{paralist}  
\usepackage{soul}

\begin{document}
\pagestyle{empty} 

\title{Gamma-Ray Burst Pulse Shapes: \\ Evidence for Embedded Shock Signatures?}

\author{Jon Hakkila\altaffilmark{1} and Robert D. Preece\altaffilmark{2}
}

\affil{$^1$Dept. Physics and Astronomy, The College of Charleston, Charleston, SC  29424-0001\\
$^2$Dept. Physics and Astronomy, University of Alabama in Huntsville, Huntsville, AL 35899\\
}
\email{hakkilaj@cofc.edu}

\section{Abstract}

A study of a set of well-isolated pulses in Long and Intermediate gamma ray burst light curves indicates that simple pulse models having smooth and monotonic pulse rise and decay regions are inadequate.  Examining the residuals of fits of pulses to such models suggests the following patterns of departure from smooth pulses:  three separate wavelike peaks found in the residuals of each pulse (the precursor peak, the central peak, and the decay peak) combine with the underlying \cite{nor05} pulse model to produce five distinct regions in the temporal evolution of each pulse. The Precursor Shelf occurs prior to or concurrent with the exponential Rapid Rise. The pulse reaches maximum intensity at the Peak Plateau, then undergoes a Rapid Decay. The decay gradually slows into an Extended Tail. Despite these distinct temporal segments, the pulses studied are almost universally characterized by hard to soft spectral evolution, arguing that the new pulse features reflect a single evolution, rather than being artifacts of pulse overlap. The fluctuations can give a single pulse the appearance of having up to three distinct localized peaks, leading to ambiguities in pulse-fitting if an incorrect pulse model is used. The approach demonstrates that complex GRBs may be composed of fewer pulses than indicated by the number of peaks. The large degree of similar spectro-temporal behavior within gamma-ray burst pulses indicates that a single process is responsible for producing pulses spanning a tremendous range of durations, luminosities, and spectral hardnesses, and the correlated characteristics of the wavelike peaks are related to the pulse asymmetry, suggesting kinematic origins that seem supportive of relativistic shocks.

\section{Introduction}

Pulses are the basic units of gamma-ray burst (GRB) emission; they are simple structures underlying the complexity of GRB light curves. Some GRB light curves are composed of a few easily-identifiable pulses while others exhibit complex light curves suggestive of many superposed pulses. In GRBs having many pulses, some emission appears to be chaotic and perhaps more rapidly variable than the pulses themselves. Although other forms of radiation are sometimes present in GRBs (extended high energy emission in a few GRBs \citep{gon03,FLAT12} and extended soft emission in some Short GRBs \citep{nor06}), the bulk of the identifiable prompt photon energy released for most GRBs is in the form of pulses. 
 
Understanding pulse behaviors is critical to explaining GRB physics, and GRB pulses exhibit many common behaviors (e.g. \cite{gol83, nor86, nor96, lia96, rrf00, nor02, nor05, ryd05, hak08}). Pulses are characterized by temporal asymmetry having longer decay rates than rise rates.  Pulses undergo hard-to-soft spectral evolution. Pulse durations are longer at lower energies than at higher energies. Pulses rise, peak, and decay faster at higher energies than at lower energies.

Many observable pulse properties are highly correlated. Shorter duration pulses tend to have brighter peak intensities, are spectrally harder, and are more time-symmetric than long duration pulses. Although there are variations among individual pulses, a statistical analysis of more than 1300 BATSE pulses \citep{hak11} demonstrates that observable pulse properties of peak flux, duration, fluence, hardness, asymmetry, and lag are significantly correlated / anti-correlated: a Spearman Rank Order correlation test gives probabilities of less than $0.01\%$ that each of these correlations are random in nature.  Pulse lag inversely correlates with pulse peak luminosity \citep{hak08}, indicating that this relation is the basis of the lag vs.\ luminosity relation obtained for integrated prompt emission \citep{nor00}.  

These correlations among pulse properties are remarkable, given that they can be easily observed in photon counts in the observer's frame, rather than, say, energy flux in the GRB emitted frames. The large cosmological distances at which GRBs are found (typically redshifts of $z \ge 1$, as established for BATSE bursts through several correlative behaviors (\cite{nor00, frr00, rei01, hak08}) and directly from subsequent observations by BeppoSAX, HETE-2, Swift, and Fermi) suggests that any intrinsic correlations should be smeared out or distorted by effects of observational cosmology (e.g. time dilation, the inverse square law, K-corrections) and instrumental effects. Rather, the processes by which and environments in which pulses originate (e.g. from relativistic ejected material traveling at Lorentz factors of $\Gamma \approx 300$) seem to be very much larger than effects resulting from universal expansion and instrumental biases.

Correlated pulse properties have been observed in a variety of GRB types and in a range of environments. They have been measured in the Long and Short burst classes in BATSE bursts (e.g. \cite{hak11}), in the Short class of Swift bursts \citep{nor11}, in GRB pulses observed by HETE-2 \citep{ari10}, in x-ray flares observed in Swift afterglows (e.g. \cite{chi10, mar10}), and in optical flares (e.g. \cite{li12}).

When GRB pulses are clearly identified as being single and isolated from other pulses, their time-resolved spectra are found to undergo hard-to-soft evolution characterized by a decay of the $E_{\rm peak}$ spectral parameter \citep{hak11}. This is consistent with the temporal delay of the pulse shape observed in low energy channels relative to high energy ones \citep{hak09, ukw10, hak11}, and also explains their asymmetric light curve shapes.

The mechanism responsible for GRB pulsed emission is still unresolved. Possible explanations include both radiative and kinematic phenomena, but no model has been able to explain all of the aforementioned observed pulse characteristics.

The standard model for GRB prompt spectral emission is kinematic energy injection into a medium via collisionless relativistic shocks; the preferred mechanism by which radiative energy is thought to be released is in the form of a synchrotron spectrum (e.g.~\cite{ree94}). However, the standard synchrotron shock model has no time-dependent component \citep{boc10}, which means that correlative pulse relations are not a direct and simple consequence of the synchrotron shock model. Synchrotron shock injections can be coupled with a cooling blackbody to explain the observations (e.g. \cite{goo86, dai02}), but they require the two energy functions to decay in tandem over the entire duration of the pulse; which is difficult, since GRB pulse durations span five orders of magnitude.

An alternate spectral evolutionary model involves Jitter radiation, which forms when electrons are accelerated by the tangled magnetic field near the shock boundary. Jitter radiation appears to produce reasonable GRB spectra (e.g. \cite{med09}). However, the temporal pulse shapes predicted from the Jitter model are very short (a fraction of a second) and have time histories which exhibit intensity-linked, rather than correlated, pulse properties such as those observed. 

The asymmetric shape of most GRB pulse light curves has led to the explanation that pulses are caused by curvature in a relativistic outflow. The basic idea is that a shell that coasts without emitting photons until it encounters another shell, then emits for a short period of time to produce a pulse rise and a decay dominated by curvature effects \citep{fen96, rrf00}. This model has been developed by many authors over the years using a variety of initial assumptions (e.g. \cite{qin02, der04,wil10}), but some parameters such as the rate of pulse decay are not found to match the observations well (e.g. \cite{koc03, pen12}).  

Recently, attempts have been made to reconcile the spectral evolution and curvature models, and these have found that both effects are needed to explain GRB pulse profiles (e.g. \cite{pen12, bas12}). However, no attempt has yet been made to place constraints on the interplay between the spectral models and the kinematic ones.

Another recent model is the Internal- Collision-induced MAgnetic Reconnection and Turbulence (ICMART) model of \cite{zha11}. This model states that collisions between relativistically expanding shells close in to the central engine serve primarily to distort magnetic field lines in the ejecta and cause no emission, while farther away these collisions provide a catalyst to allow runaway magnetic reconnections, creating pulsed emission. This model has the added value of explaining short, spiky features in GRB light curves as being due to magnetic turbulence in the emission region.

A host of empirical pulse models have also been used to fit pulse light curves; many of these have been more successful at matching the observations than the purely theoretical ones. These have included models with a power law rise and decay (e.g. \cite{nor96, lee00a, lee00b}), models with a power-law rise and exponential decay (e.g. \cite{jia05}), models with a Gaussian rise and an exponential decay (e.g. \cite{ste96}), models with a Gaussian rise and decay \citep{bha12}, and models involving more complex rise and decay functions (e.g. \cite{koc03}). A very successful empirical function is the four-parameter power-law rise and decay model of \cite{nor05}; this model has become widely used in recent years. Despite the utility of these models, their fitting parameters are not generally based on assumed physical mechanisms (an exception is that of \cite{koc03}, which has been retroactively incorporated into a theoretical scheme by \cite{zha05}), so there is no reason for choosing one over another except for goodness-of-fit in their results or the number of free model parameters.

The key to understanding the physics of GRB pulsed emission lies in finding agreement on the definition of a GRB pulse: our knowledge of GRB physics is often hindered by how investigators choose to parameterize pulse light curves. Fitting GRB pulses is generally made difficult by the low signal-to-noise regime in which they are found. The gamma-ray background is composed of discrete sources that are often themselves variable; these include the Sun, the Earth's magnetosphere, and photon scattering from inside the detecting satellites. Background rates are variable and complex: GRB experiments are located aboard satellites having eccentric orbits, and GRB instrumentation is often dominated by nonlinear response functions.  The most difficult issue affecting GRB pulse fitting, however, is the confusion caused by overlapping pulses. Many single temporal structures having the appearance of a pulse at low temporal resolution can themselves be shown to contain more than one overlapping pulse (e.g. \cite{hak11}). Thus, the signal from one pulse becomes noise to another being fitted.

When GRB pulse structures are fitted to a model (either based on a physical mechanism or empirical), how do we know whether or not a pulse-fitting technique has correctly identified the actual number of pulses? How important is the form of the fitting function to the pulse extraction process? How accurate do the fits need to be for us to incorporate better physics in our models? Most importantly, how similar or different are the intrinsic shapes of GRB pulses? Finally, how can pulse information be used to further understand the underlying physics of GRB prompt emission?  In this manuscript we will demonstrate that GRB pulse light curves exhibit distinct episodes that are common among pulses; the resulting pulse segments can be used to more clearly identify individual pulses. The fluctuations are apparently present in pulses spanning a large range of durations, intensities, and asymmetries; this information can be used to further constrain all physical pulse models.

\section{Description}
\subsection{The \cite{nor05} Pulse Model}

To register light curves with durations and amplitudes spanning orders of magnitude, we fit each light curve to the 4-parameter pulse model of \cite{nor05}: 

\begin{equation}
I(t) = A \lambda e^{[-\tau_1/(t - t_s) - (t - t_s)/\tau_2]},
\end{equation}
where $t$ is time since trigger, $A$ is the pulse amplitude, $t_s$ is the pulse start time, 
$\tau_1$ and $\tau_2$ are characteristics of the pulse rise and pulse decay, and the constant 
$\lambda = \exp{[2 (\tau_1/\tau_2)^{1/2}]}$. The pulse peak time occurs at time $\tau_{\rm peak} = t_s + \sqrt{\tau_1 \tau_2}$. 

The technique we use for our basic fitting gamma-ray burst pulses has been described elsewhere \citep{hak08, hak11}, and is summarized briefly here. Intervals in a GRB time history containing possible pulses are identified using the Bayesian Blocks routine of \cite{sca98}. Pulses are subsequently fitted using an iterative nonlinear least squares approach, with statistically insignificant pulses removed according to an algorithm that favors neither short, bright pulses nor long, fainter ones with respect to one another \cite{hak03}. Fits obtained from the summed multi-channel data are used as initial guesses for individual energy channel fits. The iteration terminates when all insignificant pulses have been removed.

Observable parameters of a fitted pulse can be defined from the four free pulse fit parameters. Measures of pulse duration $w_n$ can be defined in terms of the times when the fitted intensity has dropped to $e^{-n}$ of its maximum value. These duration  measures are obtained from the summed four-channel data by setting the intensity in Equation 1 equal to $A e^{-n}$ and solving for $t-t_s$ for the roots $(t-t_s)_+$ and $(t-t_s)_-$ are found from
\begin{equation}
2 \ \mu- \frac{\tau_1}{t-t_s} - \frac{t-t_s}{\tau_2} = -n
\end{equation}
(where $\mu = \sqrt{\tau_1/\tau_2}$) to be
\begin{equation}
(t-t_s)_{+,-} = \frac{\tau_2}{2} (n+2 \mu) \pm n \sqrt{1+4\mu /n}.
\end{equation}
The pulse decay and rise times are the positive differences between the roots and $\tau_{\rm peak}$
\begin{equation}
\tau_{\rm decay} = (t-t_s)_+ -\tau_{\rm peak} = \frac{n \tau_2}{2}[\sqrt{1+4 \mu/n} +1]
\end{equation}
\begin{equation}
\tau_{\rm rise} = \tau_{\rm peak} - (t-t_s)_- = \frac{n \tau_2}{2}[\sqrt{1+4 \mu/n} -1].
\end{equation}
The duration $w_n$ is thus
\begin{equation}
w_n = \tau_{\rm rise} + \tau_{\rm decay} =  n \tau_2 \sqrt{1+4\mu/n}.
\end{equation}
The asymmetry $\kappa_n$ is 
\begin{equation}
\kappa_n = \frac{\tau_{\rm decay} - \tau_{\rm rise}}{\tau_{\rm decay} + \tau_{\rm rise}} = \frac{n \tau_2}{w_n} = 1/\sqrt{1+4\mu/n}.
\end{equation}

We have previously defined the base pulse duration $w\equiv w_3$ as the time interval between instances at which the intensity is $e^{-3}$ ($4.98\%$) of its maximum value \citep{hak11}, or (from equation 6)
\begin{equation}
w \equiv w_3 = \tau_2 [9 + 12\mu]^{1/2}.
\end{equation}
The pulse {\em base asymmetry $\kappa$ is similarly defined as
\begin{equation}
\kappa \equiv \kappa_3 = [1 + 4\mu/3]^{-1/2};
\end{equation} }
it ranges from a value of $\kappa=0$ for a symmetric pulse to $\kappa=1$ for an asymmetric pulse with a rapid rise and slow decay. Note that this is a correction to the asymmetry definitions given in our previous pulse-fitting papers (e.g. \cite{hak08, hak11}): the published asymmetry values $\kappa_{\rm prev}$ can be approximately transformed to the new value with $\kappa \approx 0.06 \exp (2.8 \kappa_{\rm prev})$.

Equations 8 and 9 indicate that both the duration and asymmetry increase as the pulse intensity to which they are sampled decreases (e.g., increasing $n$). The asymmetry increases as $\sqrt n$ for $n << 4\mu$ and as $n$ for $n >> 4\mu$ in the interval $0 \le \kappa \le 1$, while the duration increases relative to the asymmetry as $w = \tau_2 (n+4\mu) \kappa$.  Thus, the definition of a base asymmetry (and similarly of a base duration) is needed because asymmetry is linked to the measured duration.

Pulse light curves are not as smooth as the fitting functions used to describe them. Most of the observed fluctuations represent statistical (assumed Poisson) variations in the photon counts. Of concern are unanticipated brightness fluctuations such as those due to the presence of overlapping pulses, those due to other small amplitude fluctuations that might be an un-modeled pulse component of the burst prompt emission (e.g. \cite{hak09}), or natural fluctuations that might occur if the modeled pulse shape is not an adequate representation of the true pulse shape. Any sudden, large change in pulse intensity for any reason can cause a fit to be improved by introducing additional pulses; these fitted pulses may or may not really exist.

Our pulse extraction method has two operator-selected parameters that are used to desensitize the fit to unanticipated fluctuations. These are the NCP Prior variable in the Bayesian Blocks routine (which determines sensitivity to change points in the background counts per bin) and the number of required standard deviations above the dual timescale threshold (which determines sensitivity for an acceptable pulse fit). The semi-automated nature of our approach allows the user to estimate the significance and nature of unanticipated fluctuations, under the guiding principle of Occam's razor to fit the minimum number of pulses when the presence of additional pulses cannot be firmly justified.

\subsection{The Isolated Pulse Sample}

We have been examining and fitting isolated and single pulses in a variety of gamma-ray bursts in order to better understand the severity, prevalence, and nature of unanticipated fluctuations in the GRB pulse-fitting process: our goal is to optimize the pulse fitting function, and to determine the ease with which individual GRB pulses can be uniquely extracted by an optimized fitting function (Loredo et al., in preparation). Additionally, much of the pulse database used in this study comes from an ongoing project to construct a BATSE GRB pulse catalog (Hakkila et al., in preparation). We are also applying the aforementioned pulse extraction technique to other GRB instruments; the technique has already been successfully and repeatedly applied to 64-ms, multichannel data from BATSE (\texttt{http://heasarc.gsfc.nasa.gov/docs/cgro/archive/}), Swift (\texttt{http://gcn.gsfc.nasa .gov/swift\_gnd\_ana.html}), Suzaku (\texttt{http://www.astro.isas.jaxa.jp/suzaku/HXD-WAM /WAM-GRB/grb/trig/grb\_table.html}), and Fermi's GBM experiment (Preece, 2013, private communication). 

When comparing pulses from BATSE, Suzaku, and GBM, we constrain the data to 4-channel light curves in the approximate energy range 20 keV to 1 MeV. This constraint gives us some observational consistency by allowing us to compare the observations between experiments through similar temporal and spectral windows. We will demonstrate later that self-consistent results are obtained at other energies when a larger spectral range (e.g. Fermi) is utilized.

The preliminary BATSE pulse catalog (Hakkila et al., in preparation) currently contains over 1300 pulses (mostly isolated) that have been extracted from over 600 GRBs. BATSE observed a large percentage of single-pulsed bursts: 33 of the first 100 bursts in the BATSE Current Catalog are single-pulsed, with 25 of these belonging to the Long or Intermediate \citep{hor98, muk98} classes and the remaining 8 being Short. Long single-pulsed bursts are typically faint and of moderate spectral hardness while Intermediate single-pulsed GRBs are softer, fainter, and shorter than the Long ones; BATSE's instrumental response allowed it to successfully detect many of these bursts where instruments with other characteristics (such as Swift, GBM, etc.; see \cite{ban03}) have been less successful. For example, GBM has thus far only detected a handful of unambiguously single-pulsed bursts.

In order to best carry out our study of fluctuations within pulses, we have sought to locate GRB pulses that do not overlap with other pulses, and that are found during extended time intervals where the background rate is relatively well-behaved. However, we have also sought pulses spanning a range of durations, spectral hardnesses, and asymmetries, so that we can study the interactions between unanticipated pulse fluctuations and general pulse properties. Additionally, we have desired to select pulses having sufficiently fine time resolution so that pulse fluctuations are not smeared out by the bin time. As a result, our 64 ms binned sample has precluded us from studying any traditional Short GRB pulses, although pulses from several soft, somewhat short bursts belonging to the Intermediate class \citep{hak03,hor06} have been included.

Our database currently consists of 50 pulses; three of these are from GBM and the rest are from BATSE. All data are collected in the 20 keV to 1 MeV range using 64 ms data. Most of the BATSE bursts in this database have been selected from the pulse catalog currently under development, for which the entries have not been analyzed in order of trigger number.

Table 1 contains our 50-pulse GRB sample, listing for each the Pulse ID (with asterisks denoting the Intermediate GRB pulses), the base duration $w$ and its uncertainty $\sigma_w$, the base asymmetry $\kappa$ and its uncertainty $\sigma_\kappa$, the GRB's 256 ms peak flux $p_{\rm 256}$ and its uncertainty $\sigma_{\rm p256}$, and the six pulse fit parameters ($A$, $t_s$ , $\tau_1$, $\tau_2$, the background count rate $B$, and rate of change of the background count rate $S_B$), and their formal uncertainties ($\sigma_{\rm B}$, $\sigma_{\rm S_B}$, $\sigma_{t_s}$, $\sigma_A$, $\sigma_{\tau_1}$, and $\sigma_{\tau_2}$),  and the fitted chi-square value per degree of freedom $\chi^2_\nu$ along with the associated number of degrees of freedom $\nu$. Three additional pulses, to be used as a check later, are also included at the bottom of the table.

\subsection{Chi-Square Fitting using the \cite{nor05} Pulse Model}

The \cite{nor05} pulse model fits have $\chi^2_\nu  \ge 1$, indicating that they are good but not optimal. Other models (described in the Introduction) have similar drawbacks: it is this inability of any model to completely fit the data that has led to the development of so many GRB pulse functional forms.  The \cite{nor05} models provide better fits when a large time interval is selected, as this provides more degrees of freedom for the fit, and when the signal-to-noise level is small. Overall, the \cite{nor05} model is acceptable.
However, many fits of our isolated pulse sample  deviate from high quality in non-random ways: they systematically underestimate the pulse peak and overestimate the intensity on the pulse rise. This effect is quite pronounced in the case of GRB 100707a, which was observed by both GBM and Suzaku. The fit of the GBM observations preferentially settles on a three-pulse solution, while the fit of the Suzaku observations for the same burst preferentially iterates to a two-pulse solution. Although the GBM data cannot be easily force-fitted with a two-pulse solution and the Suzaku data cannot be force-fitted with a three-pulse solution, both data sets can be fitted with a single pulse.  If the different signal-to-noise ratios of the instruments are primarily responsible for the differences, and if the single-pulse solution is the correct one, then both data sets also observe photon depletions on the pulse rise and excesses at the pulse peak, implying that a more complex pulse model is needed. Furthermore, many BATSE pulses exhibit similar depletions and excesses, at similar relative times in their light curves. In other words, {\em there is evidence that the pulse fit residuals, taken from many disparate pulses observed by different GRB experiments, show depletions and excesses of intensity at specific, similar phases of the pulse light curve. We hypothesize that the modeled pulse shape likely needs to be adjusted to account for these in-phase excesses and depletions.}

\section{Analysis}

\subsection{Differences between Observations and the Pulse Model}

To test our hypothesis that the shape of the light curve differs systematically from our model, we compare the residuals (observed minus modeled; $\rm res_i$) for the 50 light curves in our pulse sample. Since the base durations of our sampled pulses range from relatively short (1 second) to fairly long (57 seconds), simply summing the residuals will not allow a direct comparison between measurements made at similar pulse phases. Thus, we must scale each GRB pulse to its base duration, and set this base duration timescale to unity. 

The pulse amplitudes of each pulse are normalized to the maximum counts in the time interval under scrutiny. The normalized residuals for an individual pulse are thus given by

\begin{equation}
\textrm{res($t$)} = [\textrm{cts($t$)} - \textrm{model($t$)}]/\textrm{(max cts)}
\end{equation}

We recognize that pulse asymmetry might play a factor in how light curve residuals can be summed, since a pulse with a large asymmetry will peak earlier in the pulse phase (as defined by the total pulse duration) than a more symmetric one. The ``phase'' of a pulse is somewhat arbitrary, as the pulse duration is dependent on the time interval being examined (see Equation 6).  Thus, we define the  {\em fiducial time} as the time interval being examined, which we initially take to be the base duration $w$. We then define $\Phi_{i; pk}$ as the fraction of the pulse fiducial time which the pulse peak occurs in the $i^{\rm th}$ pulse, and we separately sum the residuals prior to and after the pulse peak. When recombining these residuals, we realize that each individual pulse light curve gives a slightly different value of $\Phi_{i; pk}$; values of $\Phi_{i; pk}$ range from $\Phi_{i; pk}=0.0$ for a pulse with $\kappa=1$ to $\Phi_{i; pk}=0.5$ for a pulse with $\kappa=0$. When quoting a value of $\Phi_{pk}$ for a particular sample of pulses, we refer to a mean value of $\langle \Phi_{pk} \rangle$ for the sample. 

Some pulses exhibit larger residuals than others. Since we do not want pulses with either the largest or the smallest maximum residuals to bias our results, we normalize each fitted pulse peak to a unit weight by setting $A_i=1$. A mean, scaled residual distribution is obtained by dividing the residual distribution by the number of sampled pulses.

If the differences between the modeled and actual pulses are randomly distributed in a pulse's fiducial time, then the residuals should sum to zero. If the modeled and actual pulse shapes differ systematically, then positive mean residuals (which indicate photon counts in excess of the model) and negative residuals (which indicate photon count depletions from the model) should be present. 

The residuals display trends deviating significantly from zero; Figure 1 demonstrates that there are well-defined phases (similar times relative to the pulse start, peak, and end times) during pulse light curves where the observations deviate significantly from the model. The mean residuals are plotted on a fiducial timescale corresponding to the base duration. The solid curve shows the mean residual; the sum of this and the \cite{nor05} shape is essentially a global template pulse shape.ÊThe vertical error bars (bounded by tic marks) indicate how well the template is determined (1 standard deviation).Ê The vertical lines (no tics) display the RMS range of departure from the mean shape across the population of pulses. These corrections are small, indicating that the Norris el al.\ pulse shape is a good first order approximation to the intrinsic GRB pulse shapes, but also indicating that this model can be improved upon. Figure 1 demonstrates this for the mean normalized residuals of a 48 pulse sample (the first pulse in BATSE trigger 2958 has been excluded since it was not observed with 64 ms sampling, and Fermi GRB 120326a has been excluded because we did not include data from a sufficiently long sampling time). In Figure 1, the mean modeled pulse peak time is denoted by a vertical dashed line at $\langle \Phi_{pk} \rangle=0.191$. The largest deviation (an excess at the true pulse peak, which occurs slightly after the modeled pulse peak) is about $4\%$ of the mean pulse intensity. The excess peak residuals occur because the \cite{nor05} model is too smoothly varying to fit the abrupt intensity denoting the true pulse peak, and the model predicts a lesser value of the pulse peak at a fiducial time just prior to that which is actually observed.

The mean pulse residuals exhibit several excesses overriding the smoothly varying \cite{nor05} light curve, typically followed by a depletion. The peaks of these excesses correspond to distinct times within the phase where the model differs systematically from the data. {\em These deviations from the best fit can neither be captured accurately by the \cite{nor05} fitting function nor by any other published pulse fitting function because they are not part of a model characterized by a single smooth rise and a single smooth decay.}

The excess occurring during the decay phase still appears to be increasing as the intensity drops to $A e^{-3}.$ To satisfy our curiosity as to how long the pulse intensity remains in excess of the model, we redefine the pulse duration (and asymmetry) relative to times when the intensity is $A e^{-20}$; these definitions correspond to $n=20$ in equations 6 and 7. 

Figure 2 summarizes the mean residual distribution using this $w_{20}$ duration for 28 pulses for which the background has been adequately modeled 200 seconds after the trigger. The sample size has been reduced for three notable reasons: (1) the background rates of some GRBs change in a nonlinear way on the long timescales being examined, (2) the time needed to sample several our our pulses extends beyond the end of the 64 ms data stream, and (3) two GRBs have secondary pulses that begin before the residual emission from the first pulse has ended.  The contribution occurring late in the light curve extends many times the base decay time; for some pulses this is as long as tens to (perhaps) hundreds of seconds after the trigger. This extended pulse emission quite possibly continues into the burst's afterglow phase, and, when summed over the many pulses that constitute a GRB, might be responsible for the extended GRB emission seen by \cite{con02}.  The mean pulse peak in Figure 2 is located at phase $\langle \Phi_{pk} \rangle=0.093$.

Figures 1 and 2 both hint that faint additional intensity exists prior to the beginning of the fitted pulse. To test for the existence of an early signal, we extend the start of the pulse earlier than the formal pulse start $t_s$ by an amount equal to one-tenth the pulse decay time prior to the start time $t_s$, using a decay time of $w_{8}$. In other words, since
\begin{equation}
w_{8} = 8 \tau_2 \sqrt{1 + \mu/2},
\end{equation}
the timescale $w_{\rm fiducial}$ is defined from equations 4, 5, and 10 to be
\begin{equation}
w_{\rm fid} = (\tau_{\rm pk} - t_s) + 0.1 \tau_{\rm decay} + \tau_{\rm decay} = 4.4 \tau_2 [\sqrt{1+\mu/2}+1] + \sqrt{\tau_1 \tau_2}.
\end{equation}
This fiducial timescale choice allows us to see the boundaries of the identified fluctuations in both the rise (a $1\%$ maximum excess followed by a $2\%$ maximum depletion) and decay (a $2\%$ maximum depletion followed by a $1\%$ maximum excess) portions of the pulse; this gives us a better feel for the shape of the pulse light curve. The results are shown in Figure 3, where the mean pulse start time is found at $\langle \Phi_{t_s} \rangle=0.110$ and the mean pulse peak is found at phase $\langle \Phi_{pk} \rangle=0.237$. The first faint intensity increase occurs before the main body of the pulse, just prior to the formal pulse start time $t_s$. This low intensity variation takes the form of a gradual rise, or in some bursts, a distinct bump prior to the main pulse. We call this initial feature the {\em Precursor Shelf}.

The peak of the second (and largest) excess corresponds to the main pulse peak; most of the flux is in this peak. However, as mentioned previously, the peak of the residuals occurs slightly after the fitted pulse peak, indicating that fitting function underestimates the true peak intensity and is forced to find it earlier by the rapid intensity rise immediately preceding the peak. Furthermore, the temporal structure of the pulse peak is not a ``spike'' so much as it is a falling plateau.

In general, the time histories of GRB pulses indicate that a typical pulse starts before the \cite{nor05} pulse model predicts it should, and initially brightens more slowly than the model predicts (during the {\em Precursor Shelf} phase). The intensity subsequently increases quite rapidly (during the {\em Rapid Rise} phase, but slows before reaching a brighter {\em Peak Plateau}, which is also slightly later than the model predicts. The intensity initially decreases rapidly in the {\em Rapid Decay} phase, but undergoes a re-brightening during the {\em Extended Tail} phase that keeps it brighter than the model for a significantly long time. The net effect demonstrated by the residuals is that a single pulse light curve has several components, rather than following a single smooth temporal evolution. The results demonstrate that {\em GRB pulses are {\bf not} best described as monotonically increasing and decreasing functions}.

Departures from the \cite{nor05} pulse model are generally consistent from pulse to pulse. However, there are some variations among individual pulses, which vary from pulse to pulse and from burst to burst. As a result, individual pulse shapes range from being relatively smooth to being bumpy to having two or three recognizable peaks. Some of these variations appear to be statistical in nature (arising from different signal-to-noise ratios), while others appear to be systematic (due to intrinsic differences). 

Individual pulse wave features differ systematically from one another in that the rise portions of waves are commonly compressed by various amounts relative to the decay portions; these compressions appear related to pulse asymmetry but not to other pulse properties.  This is true because the data analysis method used here explicitly recognizes pulse asymmetry by separately adding contributions to the mean residual curve from each pulse's rise and each pulse's decay. In contrast, the normalization of each pulse in duration and intensity prior to combining residuals indicates that systematic variations in pulse residual characteristics do not rely heavily on these attributes. Fluence $S$, being directly related to both the duration and the amplitude, is similarly not primarily important in explaining the residual wave compressions and expansions. 

In order to better understand how asymmetry influences more subtle characteristics of pulse shape, we have subdivided our sample into four subsets composed of (a) 12 symmetric pulses (subset ``s'') having $\kappa < 0.45$ ($\langle \kappa \rangle = 0.27$), (b) 10 asymmetric pulses (subset ``a'') having $0.45 \le \kappa < 0.67$ ($\langle \kappa \rangle = 0.58$), (b) 13 very asymmetric pulses (subset ``v'') having $0.67 \le \kappa < 0.81$ ($\langle \kappa \rangle = 0.73$), and (c) 12 extremely asymmetric pulses (subset ``x'') having $\kappa \ge 0.81$ ($\langle \kappa \rangle = 0.87$). We choose four subsets because these comprise the smallest samples in which we can still easily measure asymmetry-dependent effects: two or three subsets of 16 to 25 pulses per subset clearly demonstrate asymmetry dependence but do not have enough asymmetry bins to appropriately characterize the asymmetry dependence, whereas five or more subsets having ten or fewer pulses per subset are dominated by measurement uncertainties and individual pulse variations, even though each subset has a well-defined mean asymmetry. The mean residual waves for these four subsets are shown in Figures 4 through 7, and these panels demonstrate how the residual wave changes its own shape from being symmetric for the ``s'' and ``a'' subsets to being asymmetric for the ``v'' and ``x'' subsets.

We note that this approach to characterizing the pulse shape is not complete; the possibility exists that additional deviations from the \cite{nor05} pulse shape may be present but are undetected due to biases in our residual sampling procedure. The most likely fiducial times at which additional variations might occur are (1) as precursors beginning significantly before the main part of the pulse, (2) as variations typically shorter than the 64 ms bin size, which cannot be detected using our 64 ms resolution, and (3) very late during the pulse decay, which would not be detected because our background model does not account for long-term, nonlinear background rates (curvature), and thus is limited in how late residuals can be tracked.

Finally, we note that the \cite{nor05} pulse fitting function is not the only one that necessarily results in aligned GRB pulse residuals; it may be that the residuals of other pulse fitting functions can produce similar results. However, the wave-like pattern found in the \cite{nor05} residuals indicates that this function consistently and repeatedly oscillates between underestimating and overestimating the pulse light curve, thus ``threading the needle'' statistically and therefore always producing similarly good fits.

\subsection{An Empirical Model of the GRB Pulse Residuals}

In order to better quantify the characteristics of the residual distribution and to make a smooth residual template, we have sought to develop an empirical model to describe it. There are several caveats needed in our residual fitting approach. First, our function must be able to describe the mean residual function. Second, since each subset appears to be a time-compressed or expanded version of the same fitting function, the same functional form should be required to fit each of the subsets by only changing the fitting coefficients. Third, we use Occam's razor to favor solutions having simpler functions involving fewer fitting coefficients. Finally, we have found during the fitting process that the number of free parameters can be reduced significantly by requiring fits that are time-reversed around the residual peak, and in which the residual wave following the residual peak is an expanded version of the wave prior to the residual peak (e.g. $s f(t-t_0)=f(t_0-t)$). In order to satisfy all of these criteria, we have chosen a time-reversed functional form in which the residual wave following the residual peak is a stretched version of the wave preceding the peak:


\begin{equation}
\textrm{res($t$)} = \left\{ \begin{array}{ll}
A  J_0( \sqrt{\Omega [t_0 - t - \Delta/2]}) & \textrm{if $t < t_0 - \Delta/2$} \\
A & \textrm{if $t_0 - \Delta/2 \le t \le t_0 +\Delta/2$} \\
A  J_0(\sqrt{s \Omega [t - t_0 - \Delta/2]}) & \textrm{if $t > t_0 + \Delta/2$.}
\end{array} \right.
\end{equation}
Here, $J_0(x)$ is an integer Bessel function of the first kind, $t_0$ is the central time of the Peak Plateau, $A$ is the amplitude of the normalized Peak Plateau, $\Delta$ is the Peak Plateau's duration, $\Omega$ is the Bessel function's angular frequency, and $s$ is a scaling factor that relates the function before $t_0$ with its time-inverted form after $t_0$.

The Bessel function describes the wave appearance of the residual wave pattern reasonably well, being time-symmetric with a decreasing wave amplitude and an increasing wavelength. Furthermore, Bessel functions are solutions in cylindrical coordinates to Laplace's equation, and solutions where Bessel functions are used can involve interacting fluid flow in a cylindrical system. Solutions of this type might be applicable here, as the standard model of GRB emission involves collisionless interactions between relativistic shocks in beamed jets (e.g. \cite{geh12}). The maximum value of the residual peak occurs at the measured pulse peak (where the count rate is the highest), so the residual amplitude $A$ occurs at fiducial time $t_0$. The peak is centered $\Delta/2$ from the onset of the time-varying Bessel function.   Since no evidence indicates that the Bessel function continues beyond the third zero (following the second half-wave), we truncate the function at the third zeros $J_0 (x = \pm 8.654)$. The Bessel function is stretched by an amount $1/s$ following the pulse peak (or it can conversely be considered to be compressed by an amount $s$ preceding the pulse peak); this can be found by dividing the duration following the residual peak by the duration preceding it. The time-inverted solution effectively accounts for the fact that the wave exhibits a general symmetry around the peak plateau, and this functional choice decreases the number of free parameters needed for a good fit.

The fit to the mean pulse residuals for the 50-pulse sample is shown in Figure 8. The values of the fitting variables for the GRB pulse sample are $t_0 = 0.236 \pm 0.001$, $A=0.0349 \pm 0.0003$, $\Omega = 444 \pm 7$, $s = 0.300 \pm 0.006$, and $\Delta = 0.010 \pm 0.002$. The goodness-of-fit for the function is $\chi^2_\nu=1824$ for $\nu=995$ degrees of freedom. 

The fit to the combined data is not good. However, it describes general corrections to the pulse shape, occurring at the appropriate fiducial times, while primarily underestimating the depleted fluxes preceding and following the residual peak. The $\chi^2_\nu$ value is larger than it should be, because by averaging residuals from pulses of different duration while maintaining a constant bin size we have too strongly weighted the contributions from short pulses relative to long ones: in other words, the summed residuals no longer represent a distribution governed by purely Poisson statistics, and the sample certainly has fewer degrees of freedom than the value given. More importantly, in constructing this sample template we have merged residuals from pulses spanning a large range of asymmetry values, and have thus assured that the contributions from the pulse start, pulse end, and pulse peak will all be in phase while simultaneously ignoring the possibility that contributions from the pulse rise or pulse decay may not be in phase. We will show below that asymmetry-dependent fits resolve this problem: the general form of the fitting function provides much better fits to asymmetry-dependent subsamples and to individual pulses than it does to the overall sample. Thus, the fitting function is generally a good representation of individual GRB pulse residuals, even though it does not provide a great fit to the overall dataset from which it was derived.

The fitting coefficients can be combined with the fitting function to construct a template for correcting the \cite{nor05} pulse shape, and thus for obtaining a more adequate representation of a GRB pulse's light curve. Although the template is well-characterized for the sample, the coefficients are not adequate for fitting individual GRB pulses. This is because each pulse shape has a specific asymmetry $\kappa$, and the characteristics of the mean GRB pulse residual distribution are sensitive to the mean pulse asymmetry $\langle \kappa \rangle$. Estimates of the $\kappa$-dependence of the fitting coefficients $t_0$, $A$, $\Omega$, $\Delta$, and $s$ can be obtained by separately fitting the ``s,'' ``a'', ``v'', and ``x'' subset templates.  

Our results are simplified by recognition that the parameter $\Delta$ is relatively constant across the sample. Therefore, we set $\Delta=0.01$ for all subsets, and reduce our model from a five-parameter model to a four-parameter one. The four-parameter model is a better fit to the subsets than it is to the combined residual curve, with reduced $\chi^2$ values of $\chi^2_s=1.0$, $\chi^2_a=1.8$, $\chi^2_v=1.7$, and $\chi^2_x=1.2$ for the ``s,'' ``a,'' ``v,'' and ``x'' curves, respectively (given 961 degrees of freedom. Again, we present the caveat that these values are larger than they should be, arising as they do from summing residuals from pulses with different binning.





The values of the subset template fitting coefficients $t_0$, $A$,  $\Omega$, and $s$ are given in Table 2 and are plotted as large diamonds in Figures 9 through 12 relative to the mean pulse asymmetry. General correlations are found to exist between the coefficients and pulse asymmetry. First, $t_0$ (the time of the true pulse peak measured relative to the fiducial timescale) occurs later for symmetric pulses than it does for asymmetric pulses, because symmetric pulses have longer rise times than asymmetric pulses. Second, asymmetric pulses tend to have slightly larger residual peak amplitudes than symmetric pulses. This indicates that the peak plateau is brighter for asymmetric pulses than it is for symmetric pulses, making the \cite{nor05} model a poorer fit for asymmetric pulses. Third, asymmetric pulses have larger $\Omega$ values than symmetric pulses, indicating that Peak Plateau drops off more rapidly relative to the fiducial timescale. Finally, the residual wave feature expands after the pulse peak, as indicated by the $\kappa-$dependence of $s$.

The $s$-dependence is important in the pulse's evolution, because it  shows that {\em the temporal features during the pulse decay have memory of the temporal features on the pulse rise, even though the pulse rise and decay have traditionally been considered to be unrelated pulse components.} The amount of stretching is larger for asymmetric pulses than it is for symmetric pulses, just as the ratio of decay time to rise time is larger for asymmetric pulses than it is for symmetric pulses. 

It is important to remember that the residual curve after the pulse peak is best described by a {\em time-reversed} function from before the pulse peak. This model is successful even when applied to pulses having different asymmetries, suggesting that the pulse fluctuations occurring prior to the peak time are indeed mirrored during the pulse decay, and stretched also. {\em This mirroring and stretching effect is extremely suggestive of forward- and reverse-shocked emission.}

The template approach is successful in matching the co-added observations of GRB pulse residuals. However, our interest lies in applying the technique to fits of individual pulse light curves, rather than ensembles. Even though the approach may eventually provide insights into the physical mechanism that produces the pulsed emission, it is currently one of empirical statistical pulse fitting rather than theoretical modeling, and we remind the reader that this approach still has limitations. For example, we note that the combined pulse structure (\cite{nor05} model plus residual model) might equally well be fitted by two or three separate \citep{nor05} pulse shapes, provided these shapes are temporally linked with one another to form a more complex structure. The correlated properties of the template, however, indicated that the timing of pulses in such a model must be tied to pulse asymmetry and must likely display some temporal mirroring in order to satisfy the observations.

\subsection{Modeling the Light Curves of Individual GRB Pulses}

We apply the template approach to individual pulses residuals, using the \cite{nor05} properties of the 50 BATSE and GBM pulses extracted in Table 1. The pulse's asymmetry $\kappa$ can be used with the asymmetry-dependent templates to provide initial guesses as to the template coefficient values. The process is essentially identical to that used previously to obtain the templates, although the residual coefficients have been transformed from the fiducial timescale to the actual timescale via the relations

\begin{equation}
t_0=t_{0; \rm fiducial} (t_{\rm end}-t_{\rm start}) +t_{\rm start},
\end{equation}
\begin{equation}
\Omega=\Omega_{\rm fiducial}/(t_{\rm end}-t_{\rm start}),
\end{equation}
and 
\begin{equation}
\Delta=\Delta_{\rm fiducial} (t_{\rm end}-t_{\rm start}),
\end{equation}
where $t_{\rm start}$ and $t_{\rm end}$ define the fiducial window as described in Equation 12.

The results of the individual pulse residual fits are shown in Table 3 and Table 4. Table 3 contains the coefficients  $t_0$, $A$, $\Omega$, and $s$, along with their uncertainties $\sigma_{t_0}$, $\sigma_A$, $\sigma_\Omega$, and $\sigma_s$ for each fitted pulse. Table 4 indicates the qualities of each fit, by listing $t_{\rm start}$ and $t_{\rm end}$ used for each fit, the quality $\chi^2_{\rm p}$ of the \cite{nor05} pulse fit in this interval, the number of degrees of freedom $\nu_{\rm p}$, and reduced chi-square value $\chi^2_{\nu; \rm p}$ for each fit. These results are accompanied by the quality of the fit to the pulse model plus the fit to the residual wave $\chi^2_{\rm p+r}$ in the same time interval, with $\nu_{\rm p+r}$ degrees of freedom, for a reduces chi-square value of $\chi^2_{\nu; \rm p+r}$.

Comparison of the $\chi^2_{\nu; \rm p}$ to $\chi^2_{\rm p+r}$ values in Table 3 demonstrates that fitting the residuals generally provides an improvement over fitting the pulse model alone. Most of the cases showing no improvement are those for which the $\chi^2_\nu$ values are small in both cases, indicating that the data have been over-characterized. In these cases, either the pulse is short or it has a significant period during which 64 ms resolution was not available. These results demonstrate that our empirical pulse shape fits light curve data quite well, if not perfectly. Since we do not have a physical model for the correct mechanism for GRB prompt emission, undoubtably our empirical shape is not the correct one. However, the proposed one is {\em better} from a statistical standpoint than the smooth model of Norris.

The coefficients obtained for all 50 individual pulses (along with the five test pulses) are compared to the coefficients of the templates from which they have been derived in Figures 9 through 12. The individual pulses show general correlations similar to those obtained for the templates: three of the four coefficients correlate strongly with $\kappa$, as indicated by a Spearman rank order correlation test. The correlations are (a) the probability that the anti-correlation of $-0.928$ between $t_0$ and $\kappa$ is random is $p_{t0}=1.8\times10^{-27}$, (b)  the probability that the correlation of $0.218$ between $A$ and $\kappa$ is random is $p_A=0.12$,  the probability that the correlation of $0.474$ between $\Omega$ and $\kappa$ is random is $p_{\Omega}=3.3\times10^{-4}$, and  the probability that the anti-correlation of $-0.866$ between $s$ and $\kappa$ is random is $p_s=5.7\times10^{-17}$.

The individual pulse $t_0$ distribution deviates slightly from the template coefficient distribution in that all pulses with $\kappa \ge 0.5$ all have larger $t_0$ values than the corresponding template values (we exclude the three test pulses having similarly small $t_0$ values, since they were not used to construct the templates). The reason for this is that template residuals sum together emission from a dozen pulses, so each subset template has a longer Extended Tail than the individual pulse measurements. A longer Extended Tail increases $t_{\rm end}$, which has the effect of moving $t_0$ back to an earlier time relative to $t_{\rm end}$. The effect is less noticeable for symmetric pulses, where early emission also moves $t_{\rm start}$ to an earlier time, and this keeps $t_0$ in the same place relative to the fiducial timescale. Additionally, pulses taken from detectors with smaller surface areas (such as GBM) have smaller $t_0$ values because they observe less of the pulse tail emission.

We present three examples of the resulting pulse fits: BATSE triggers 3026 (Figure 13), 3040 (Figure 14), and 469 (Figure 15). The fits to the residual model are shown in the left-hand panels, while the summed fit of the pulse model plus the residual model are shown in the right-hand panels. BATSE trigger 3026 is a somewhat symmetric pulse ($\kappa = 0.56$), and the residual pattern is correspondingly symmetric ($s = 0.36$), while BATSE trigger 3040 is more asymmetric ($\kappa = 0.91$) with a correspondingly more asymmetric residual pattern ($s=0.07$).  For these pulses, the residual waves are pronounced and easily observed, and the total fits to the light curve accurately accounts for the observed undulations and fluctuations. Figure 15 demonstrates the fit for BATSE trigger 469, which was not used in the original analysis. The fit to this pulse will be described below.

The residual fits and combined pulse fits describing the corrected pulse shapes are also demonstrated in color for each of the 50 sampled pulses in the electronic journal (the first two panels of  online Figures 20 through 69), as well as for five other pulses used to test the results (the first two panels of online Figures 70 through 74). The results indicate that {\em many apparently very different pulse light curves can now be simply explained by the way in which the basic pulse shape is augmented by the asymmetry-dependent residual curve.} Because the residual wave alignment depends on pulse asymmetry, the resulting pulse shape (single-pulse model plus residuals) can produce pulses which look different from one another, but which represent the same general underlying evolution. 

\subsection{What is the Definition of a GRB Pulse? Three {\bf Isolated} Test Pulses}

The pulse template results indicate that pulse light curves are often characterized by up to three monotonic intensity increases, rather than a single one, and our analysis demonstrates that the fiducial times at which the secondary and tertiary (reflected) bumps occur are predictable from the pulse asymmetry. Thus we hypothesize that there might be GRB light curves exhibiting several distinct bumps that could be more simply explained by a single-pulse model than by a multiple-pulse model. 

To test this hypothesis, we have selected three GRBs having isolated emission episodes that appear to contain more than a single pulse (BATSE 0469, BATSE 3164, and the second emission episode of BATSE 2958). We have forced the pulse-fitting routine to fit each of these with only a single pulse. We can then examine the light curve residuals to see if they can be explained by the template results.

BATSE trigger 469 (Figure 15) is one of the three test pulses not originally used in the calibration of the residual distribution. The semi-automated pulse-fitting code had favored a three-pulse solution for this burst, but we forced a single-pulse \cite{nor05} pulse solution and verified that the deviations from the pulse model aligned with our asymmetry-dependent residual template. This moderately asymmetric pulse ($\kappa = 0.80$) indeed also has a correspondingly asymmetric residual wave pattern ($s=0.21$).

The residual curves and fitting coefficients of the three pulses are indeed consistent with those used in developing the template model; the residual curves have similar shapes and occur at similar fiducial times to the template pulses. The main difference is that the side lobe amplitudes of the residual curves are much larger than those of the template pulses. Since BATSE 0469, BATSE 3164, and BATSE 2958 originate in bright bursts, it seems possible that these pulses represent better examples of the template shape than the pulses used to construct the templates. If we had included these three pulses in the construction of our templates, then that predicted amplitudes of our light curve bumps would be larger than they are now. {\em The residual curves of these bright pulses further supports the idea that a standard GRB pulse light curve consists of three distinct bumps.}

\subsection{Spectral Evolutionary Support of the Single-Pulse Interpretation: the Role of the Precursor Shelf}

The Precursor Shelf appears to be an important signature of the pulse energy release process. As such, the spectral characteristics of these separate temporal features are important. Pulse spectral evolution has been generally described as a hard- to soft- process, although some observers have previously measured some pulse hardnesses that track with intensity. Hakkila and Preece (2011) have demonstrated that most identified intensity tracking pulses are found in complex GRBs, where many overlapping pulses are present. Intensity tracking in most of these cases can be explained by overlapping hard- to- soft pulse evolution that resets quickly to hard when a trailing pulse begins. However, there may be isolated instances where intensity tracking occurs.

Since we now recognize the existence of phased intensity fluctuations in a GRB pulse light curve, we perform some simple tests of its spectral hardness relative to the emission of the rest of the pulse. A simple measure of hardness in low signal-to-noise regimes can be made by dividing the 100 keV to 1 MeV emission (BATSE channels $3+4$) by the 20 to 100 keV emission (BATSE channels $1+2$). This counts hardness HR(12/34) can be studied as a function of time in the sampled GRB pulses. 

Figure 16 presents the hardness evolution for a sample of ten representative GRB pulses (the third panel of online Figures 20 through 72 demonstrate the hardness evolution of 53 of the 55 GRB pulses used in this analysis). The first time bin for each pulse generally contains emission prior to the burst trigger, including emission from the Precursor Shelf and/or part of the pulse rise. This might not have previously been considered to be pulse emission, but its presence supports the idea that the Precursor Shelf is a critically important part of the evolving pulse. The formal pulse decay, occurring after the pulse peak, represents a continued, monotonic decrease in pulse hardness begun during the intensity rise. The second and third time bins for each pulse typically contain the hardness during the rapid pulse rise; the pulse hardness continues its decline during these times. Each pulse subsequently softens until the hardness in these channels can no longer be measured, and the pulse harness measure ``ends'' when the high energy signal is lost, suggesting that it continues emitting at lower energies at later times, after the process has finished producing higher energy photons. 

Pulses having the shortest durations represent the most rapid spectral decays, supporting the idea (\cite{hak09,hak11,pen12}) that a pulse is generally {\em defined} by its hard-to-soft evolution. Furthermore, short spiky pulses (primarily those from Short GRBs) and asymmetric pulses also appear to begin with harder emission than long, symmetric pulses; short and asymmetric pulses begin with a greater percentage of hard photons and thus evolve more than long symmetric pulses.

We seek to verify that this evolution is not simply a characteristic of isolated pulses, and that it is not observed only in BATSE data, by studying the $E_{\rm peak}$ evolution of two pulses found in a bright, complex GBM burst. GRB 130427a is a nearby energetic burst \citep{pre13} having a complex $\gamma$-ray light curve exhibiting many peaks. The brightest part of the light curve is chaotic, consisting of multiple hard spikes. This emission is followed by smoother fluctuations, and finally ends with an isolated pulse. The time of the trigger appears to be characterized by one or more distinct pulses. 

We have chosen to force-fit the emission earlier than 2.6 seconds after the trigger with a single pulse, then examine the residuals to see if they are consistent with the residual curves found for the other GRB pulses in this sample. This pulse, considered alone, is brighter than all but four GRBs observed by GBM so it could be analyzed both spectrally and temporally with unprecedented detail. The pulse width in seconds is well described as a power law as a function of energy and the pulse lag as function of energy follows this same model without any modification. As such, we force-fit because a single pulse is not an optimal solution. We improve the fit by also fitting the residuals in the manner described previously. Here, we expect that pulse residual wave deviations from the calibration characteristics will be amplified. Indeed, $\chi^2_\nu = 16.15$ for 46 degrees of freedom for the pulse plus residuals fit, but this is an improvement over the fit for a single pulse alone ($\chi^2_\nu = 23.33$ for 50 degrees of freedom). Furthermore, the residual curve characteristics found during the fiducial timescale (Figure 17), summarized in Tables 1 and 3, are consistent with those of other isolated pulses (Figures 9 through 12). Although it is found in the complex, overlapping light curve of a bright Fermi burst, the trigger emission appears entirely consistent with emission from a single pulse rather than from several overlapping pulses. 

The final, isolated pulse in GRB 130427a behaves in a manner similar to the calibration pulses. Figure 18 demonstrates the pulse fit, the results of which are summarized in Tables 1 and 3. However, we note that the background rate at the beginning of the pulse is higher than it was before the trigger pulse and is declining when the final pulse begins, suggesting that the previous emission episode had not entirely ended.

We examine the spectral behavior of these two pulses in GRB 130427a by viewing the decay of $E_{\rm peak}$ taken from both GBM and LAT data, rather than by examining a hardness ratio. The results, shown in Figure 19, demonstrate that the hardness evolution seen in isolated BATSE pulses represents a more extensive and profound spectral decay than can be summarized by a simple hardness ratio evolution, which is consistent with the interpretation of $E_{\rm peak}$ decay found by \cite{pen12}. The hardest part of the trigger pulse, as defined by $E_{\rm peak}$, occurs during the Precursor Shelf. The same would be true for the Precursor Shelf of the final pulse, except that the soft tail of the previous emission episode seems to soften the emission from the Precursor Shelf somewhat. The increase in $E_{\rm peak}$ at the end of the pulse is difficult to measure, but coincides with the onset of the afterglow and thus may be affecting the pulse measurements. In general, the spectral evolution observed here is hard to soft and clearly supports the interpretation that spectral decay is a defining characteristic of GRB pulsed emission.

{\em The residual intensity fluctuations observed here seem to be independent of the pulse's spectral evolution, e.g. they do not significantly change or alter the GRB pulse spectral decay process. This suggests that the mechanism responsible for pulse spectral evolution is a stronger and more dominant source of photons than the weaker and presumably kinematic signature originating in the residual distribution.} 

In summary, the Long and Intermediate GRB pulses observed in this study generally exhibit hard-to-soft evolution beginning with a faint initial emission bump or peak. This fluctuation is indeed part of the pulse, and it may be a signature of the initial energy injection that initiates the subsequent pulse spectral decay process. The pulse energy continues to decay, even as the intensity rises again on the bump or peak that occurs on the decay portion of the light curve. The correlated pulse properties found in both Long and Short GRB classes (e.g. \cite{hak11}) suggest that Short burst pulses have similar spectral evolution and might therefore have similar temporal characteristics as well, but our choice of 64 ms data does not provide us with sufficient temporal resolution to disprove or verify this hypothesis.

\section{Discussion}

Individual GRB pulse light curves clearly depart from smooth and monotonic pulse rise and decay. In Long and Intermediate GRB pulses, the deviations from a smoothly varying monotonic pulse rise and decay take the form of a low amplitude wavelike residual structure that is time-symmetric, with a symmetry axis near the pulse peak. The residuals can be modeled using a similar functional form for all GRB pulses, although the wave structure occurs at different fiducial timescales and has different amplitudes from pulse to pulse. The coefficients describing the wave properties correlate with the pulse asymmetry, but not with the pulse duration. The residual waves appear to have at least three well-defined peaks: the {\em precursor peak} occurs either at or before the pulse rise; it can be observed as a separate faint pulse in pulses that are symmetric or as a shelf on the rise of pulses that are asymmetric. In general, the precursor peak is spectrally the hardest part of a pulse. The {\em central peak} often coincides with the peak of the pulse itself, and is better described as a plateau than as a sharp peak. Although the central peak indicates the time at which pulse is brightest in the detector channels, it is not {\em spectrally} unique, as the pulse hardness continues its fall through the peak from its maximum value at the onset of the precursor phase. However, the existence of the peak plateau structure argues that the pulse peak is {\em temporally} unique. It does not always appear to be directly aligned with the fitted pulse peak: the central peak often lags behind the pulse centroid, and, in the case of the second pulse of BATSE trigger 2958, clearly leads it. The {\em decay peak} occurs on the decay and constitutes a large portion of the decay phase. As a result of this peak, the decay rate declines and in some cases also rises for a while before decaying. The decay peak appears to be a stretched, time-reversed version of the precursor peak, although it is much softer than either the precursor pulse or the central peak. It is temporally similar to yet spectrally different from the precursor peak.

The shortest timescale on which the residual pulse variations occur is typically equal to the pulse rise time (for asymmetric pulses) or shorter than it (for symmetric pulses). Thus, the wave features that appear during the precursor pulse are typically the shortest timescale features observed in a pulse. The wavelike variations perhaps explain why some GRBs are observed to have minimum timescale variabilities significantly shorter than their pulse rise times \citep{mac12}.

The theoretical focus in the GRB literature has been on studying and explaining monotonically increasing and decreasing single-peaked pulses; this is perhaps natural because the evidence previously indicated a single-peaked pulse nature. However, some theoretical studies have led to the possibility that pulses are multi-pulsed. For example, \cite{kin04} have used a simple relativistic model to demonstrate that three peaks should occur in the case of two colliding shells under the specific condition when the shells have equal bulk kinetic energies. In order to have equal energies, the slower shell has a greater mass than the more rapidly moving shell catching up with it. \cite{kin04} predict that three density peaks are produced by forward and reverse rarefaction waves moving through the shells and across the contact discontinuity. However, these authors also predict that the short timescale of the interaction between the shocked regions causes the dips between the peaks to be smeared out, effectively making the pulse single-peaked. Their analysis does not yet describe the way in which density structures eventually contribute to the production of GRB spectral emission. However, their explanation is consistent with these observations and might be pursued in greater detail, to see if it is in agreement with observed pulse residual characteristics.

The pulse residual patterns seen here demonstrate intensity variations that are not directly related to the pulse hardness evolution. They therefore seem to provide, perhaps for the first time, an indication that every GRB pulse contains a kinematic signature which accompanies the more well-known spectral decay. The close correlation between pulse asymmetry $\kappa$ (indicating the rate of hard-to-soft photon energy emission), the success of a time-reversed function in characterizing the residual fluctuations, and the stretching of the wavy component $s$ (which we hypothesize to be a kinematic signature of the shell collision) unites these two. From the perspective of a physical model, faster relative shell velocities should lead to shorter impact times, and to higher amplitude, more asymmetric pulses having more asymmetric wavy patterns.

\section{Conclusions}

We have uncovered a non-monotonically varying component demonstrating that GRB pulses spanning orders of magnitude in duration and amplitude have very similar shapes after a simple pre- and post-peak temporal rescaling. The varying component appears to be present in \st{all} {\bf many} GRB pulses, and provides evidence that GRB pulses exhibit more complex structure that is related to the asymmetry of the pulsed emission. 

The varying component can be fitted with a three-peaked wavelike structure overlapping the monotonic pulse rise and decay, after an underlying \cite{nor05} pulse shape has been subtracted. The entire pulse light curve, including this wavelike structure, undergoes a natural hard-to-soft evolution. The wavelike structure has three peaks. The middle or central peak corresponds closely with the position of the pulse peak; this peak appears to be a short-duration plateau rather than a sharp peak. The precursor peak occurs coincident with or prior to the pulse rise; this peak is observed as a faint hard pulse or as a shelf on the pulse rise.  The decay peak occurs during the decay; it seems to be a soft, stretched, time-reversed version of the precursor peak which increases the pulse's duration.

The fact that GRB pulse light curves generally exhibit these phased residual components makes it easier to identify and isolate individual pulses within a GRB light curve. These undulating light curve shapes help remove our preconceptions that individual GRB pulses must have monotonically increasing and decreasing light curves. If a light curve exhibits phased undulations similar to those shown here, then it is much more likely to be a single pulse than one with undulations that are not phased appropriately.

The data do not allow us to discern at present whether the pulse shape variability is due to variability of a single pulse structure, or whether it is due to overlapping independent pulse structures. Regardless of which model is correct, the emission following the pulse peak is linked to the emission preceding the pulse peak, even though the spectral hardness decreases throughout the pulse.

These findings have important implications for phenomenological and physical modeling of GRBs.  The specific features we find potentially constrain such models in several ways:

\begin{itemize}
\item Physical models must account for the approximate uniformity of pulse shapes.
\item Although the flux rises and falls dramatically over the course of a pulse, the hardness decreases nearly monotonically; physical models must account for this surprising spectro-temporal behavior.
\item Physical models must explain why pulse asymmetry also leads to asymmetry in the template pattern.
\item Physical models must explain why properties of the smooth pulse emission such as duration, hardness, intensity, fluence, and asymmetry all correlate, and are thus also correlated with the template properties.
\item The correlated pulse properties found in both Long and Short GRBs (e.g. \cite{hak11}) suggest that Short GRB pulses have similar spectral evolution and might therefore have similar temporal characteristics; higher time resolution analysis is needed to explore this.
\end{itemize}

This exploratory analysis in the spirit of {\em functional data analysis} (FDA) methods developed in statistics, and our work points to the potential value of FDA for GRB pulse studies. These results are guiding our ongoing GRB pulse analyses. Better application of the techniques developed here may eventually lead to extraction of pulses in complex bursts containing many overlapping pulses. Some pulsed GRB emissions have not yet been demonstrated to exhibit these scaled shapes; these emissions include Short burst pulses, spiky emission emanating from complex Long bursts, and extended emission found in Short bursts. Pulses can be less ambiguously identified and measured in GRB light curves if a corrected pulse model, accounting for the residuals, is used instead of an idealized model. Furthermore, application of new techniques such as L\'{e}vy Adaptive Regressive Kernals (e.g. \cite{wol11}) will allow for more accurate extraction of overlapping pulses (Broadbent et al.\ 2014, in preparation).

The separate peaks in each single GRB pulse suggest evidence of kinematics consistent with collisions between shells in a relativistic outflow. The presence of multiple peaks, coupled with continuous energy decay, may be intrinsic signatures embedded within every GRB pulse. The mirrored, stretched characteristics of the intensity fluctuations in the pulse residuals correlate with pulse asymmetry and are suggestive of forward and reverse relativistic shocks, and/or of interactions in a compressed fluid. However, the relationship between the mechanism responsible for the pulsed radiation and kinematic signatures is still unknown. Future studies involving the information embedded in GRB pulse shapes can lead to better models and can therefore help resolve these issues.

\section{Acknowledgements}
Without the assistance and insights of Thomas Loredo, Robert L. Wolpert, and Mary E. Broadbent, it is safe to say that this manuscript would have turned out very different than it did. Although these collaborators opined that their contributions did not warrant co-authorship, we cannot but strongly disagree: in this age of hundred-author papers, their contributions have been of profound importance, and we gratefully thank them. We acknowledge valuable theoretical discussions with Demos Kazanas and Matthew Baring, and useful general discussions with Kalvir Dhuga and Eda Sonbas. Finally, we acknowledge the helpful comments from the anonymous referee, which greatly improved this manuscript. This work has been supported by NASA ROSES AISR program grant NNX09AK60G and from GBM through NNM11AA01A/MSFC.

\clearpage

\begin{deluxetable}{ccrrrrrrrrrrrrrrrrcrl}
\tabletypesize{\scriptsize}
\rotate
\tablecaption{GRB pulses used in this analysis.\label{tbl-1}}
\tablewidth{0pt}
\tablehead{
\colhead{Pulse ID} & \colhead{$w$} & \colhead{$\sigma_w$} & \colhead{$\kappa$} &
\colhead{$\sigma_\kappa$} & \colhead{$p_{\rm 256}$} & \colhead{$\sigma_{\rm p256}$} &
\colhead{$B$} & \colhead{$\sigma_{\rm B}$} & \colhead{$S_B$} & \colhead{$\sigma_{\rm S_B}$} &
\colhead{$t_s$} & \colhead{$\sigma_{t_s}$} & \colhead{$A$} & \colhead{$\sigma_A$} &
\colhead{$\tau_1$} & \colhead{$\sigma_{\tau_1}$} & \colhead{$\tau_2$} & \colhead{$\sigma_{\tau_2}$} &
\colhead{$\chi^2_\nu$} & \colhead{$\nu$}
}
\startdata
BATSEÊ 332 & 36.3 & 0.8 & 0.80 & 0.01 & 1.188 & 0.110 & 503.3 & 0.1 & -0.016 & 0.002 & -1.30 & 0.09 & 264.8 & 2.9 & 1.70 & 0.18 & 9.71 & 0.24 & 0.8835 & 7501 \\
BATSEÊ 493* & 4.5 & 0.5 & 0.39 & 0.03 & 1.096 & 0.121 & 485.4 & 0.3 & 0.112 & 0.002 & -1.90 & 0.35 & 245.0 & 6.6 & 10.52 & 4.65 & 0.58 & 0.08 & 0.8117 & 7501 \\
BATSEÊ 501 & 9.4 & 0.5 & 0.82 & 0.07 & 1.099 & 0.118 & 472.9 & 0.3 & -0.067 & 0.002 & -0.87 & 0.05 & 196.9 & 5.4 & 0.32 & 0.09 & 2.58 & 0.15 & 0.8228 & 7469 \\
BATSEÊ 540 & 3.8 & 0.3 & 0.79 & 0.11 & 0.759 & 0.108 & 1004.7 & 0.5 & 0.037 & 0.002 & -0.71 & 0.03 & 319.1 & 11.5 & 0.21 & 0.07 & 1.01 & 0.08 & 0.9219 & 7501 \\
BATSEÊ 563 & 30.0 & 0.6 & 0.87 & 0.02 & 1.886 & 0.143 & 758.9 & 0.5 & -0.130 & 0.002 & -0.75 & 0.04 & 389.6 & 4.2 & 0.48 & 0.05 & 8.75 & 0.17 & 0.9181 & 7465 \\
BATSEÊ 658 & 18.2 & 0.8 & 0.64 & 0.02 & 1.281 & 0.149 & 430.1 & 0.8 & 0.099 & 0.014 & -4.13 & 0.22 & 204.7 & 3.6 & 4.63 & 0.85 & 3.88 & 0.21 & 1.1540 & 2024 \\
BATSE  673 & 43.7 & 17.1 & 0.18 & 0.01 & 0.299 & 0.105 & 496.9 & 0.7 & 0.054 & 0.008 & -51.73 & 28.90 & 46.5 & 1.8 & 1323.28 & 1963.10 & 2.62 & 1.29 & 0.9811 & 3118 \\
BATSE  680 & 9.5 & 2.2 & 0.19 & 0.02 & 0.814 & 0.106 & 723.1 & 0.4 & -0.043 & 0.002 & -10.34 & 3.48 & 192.1 & 5.0 & 232.50 & 205.90 & 0.61 & 0.18 & 0.8148 & 7501 \\
BATSE  711*$\dagger$ & 1.9 & 0.3 & 0.28 & 0.04 & 1.048 & 0.147 & 745.4 & 0.4 & -0.465 & 0.002 & -1.50 & 0.36 & 395.4 & 12.5 & 13.73 & 9.71 & 0.18 & 0.04 & 0.8598 & 7501 \\
BATSE  727*$\dagger$ & 6.0 & 1.0 & 0.31 & 0.03 & 0.874 & 0.116 & 844.5 & 0.4 & -0.008 & 0.002 & -3.82 & 0.93 & 218.7 & 6.9 & 29.80 & 19.74 & 0.63 & 0.13 & 0.8084 & 7469 \\
BATSEÊ 795 & 21.0 & 1.2 & 0.76 & 0.02 & 1.115 & 0.129 & 484.6 & 0.6 & -0.052 & 0.001 & -1.76 & 0.17 & 169.8 & 3.3 & 2.10 & 0.40 & 6.78 & 0.32 & 1.19 & 3118 \\
BATSEÊ 907 & 18.3 & 0.4 & 0.83 & 0.02 & 3.569 & 0.169 & 489.5 & 1.2 & 0.095 & 0.042 & -0.39 & 0.03 & 591.2 & 6.0 & 0.61 & 0.06 & 5.03 & 0.10 & 1.52 & 1087 \\
BATSEÊ 914 & 6.1 & 0.2 & 0.65 & 0.03 & 2.533 & 0.163 & 502.1 & 1.0 & 0.183 & 0.036 & -0.86 & 0.06 & 435.0 & 7.2 & 1.45 & 0.24 & 1.32 & 0.06 & 1.109 & 1087 \\
BATSE 1039 & 18.9 & 1.1 & 0.40 & 0.01 & 1.383 & 0.117 & 549.3 & 0.4 & -0.035 & 0.002 & -5.74 & 0.72 & 238.0 & 3.4 & 40.20 & 9.13 & 2.50 & 0.18 & 0.8406 & 7465 \\
BATSE 1145 & 3.4 & 0.3 & 0.48 & 0.04 & 1.762 & 0.148 & 510.5 & 0.6 & -0.033 & 0.007 & -1.09 & 0.16 & 301.9 & 8.6 & 3.53 & 1.34 & 0.54 & 0.06 & 1.13 & 3118 \\
BATSE 1200 & 23.5 & 1.4 & 0.33 & 0.01 & 1.169 & 0.117 & 547.8 & 0.4 & -0.029 & 0.002 & -10.96 & 1.21 & 248.2 & 3.0 & 99.50 & 23.20 & 2.57 & 0.19 & 0.8263 & 7469 \\
BATSE 1301$\dagger$ & 25.6 & 13.5 & 0.20 & 0.03 & 0.360 & 0.108 & 554.5 & 0.4 & 0.122 & 0.005 & -31.60 & 20.71 & 44.3 & 2.6 & 586.26 & 1173.88 & 1.67 & 1.11 & 1.051 & 3903 \\
BATSE 1306 & 10.8 & 3.0 & 0.30 & 0.03 & 0.542 & 0.117 & 826.3 & 0.8 & -0.235 & 0.027 & -7.77 & 2.81 & 106.7 & 5.2 & 59.80 & 63.86 & 1.09 & 0.38 & 1.12 & 1556 \\
BATSE 1319*$\dagger$ & 8.6 & 1.0 & 0.66 & 0.07 & 0.733 & 0.125 & 799.4 & 0.8 & -0.156 & 0.015 & -2.71 & 0.25 & 144.4 & 6.5 & 1.82 & 0.89 & 1.89 & 0.26 & 1.162 & 2341 \\
BATSE 1379*$\dagger$ & 2.8 & 0.3 & 0.57 & 0.08 & 0.463 & 0.111 & 494.8 & 1.5 & 0.035 & 0.244 & -1.00 & 0.12 & 237.3 & 9.9 & 1.29 & 0.62 & 0.54 & 0.08 & 1.235 & 310 \\
BATSE 1406 & 26.8 & 0.5 & 0.75 & 0.01 & 1.968 & 0.133 & 483.3 & 0.7 & 0.026 & 0.008 & -1.47 & 0.08 & 431.3 & 3.6 & 2.58 & 0.20 & 7.25 & 0.14 & 1.185 & 3118 \\
BATSE 1432 & 57.3 & 6.9 & 0.57 & 0.01 & 0.503 & 0.109 & 792.1 & 1.5 & -0.206 & 0.043 & -16.09 & 1.78 & 92.0 & 2.5 & 27.33 & 10.22 & 10.81 & 1.54 & 1.124 & 1560 \\
BATSE 1446 & 30.5 & 12.4 & 0.11 & 0.01 & 0.483 & 0.111 & 552.2 & 0.4 & 0.072 & 0.002 & -62.88 & 34.77 & 91.1 & 2.4 & 4084.55 & 6301.33 & 1.13 & 0.56 & 0.8337 & 7497 \\
BATSE 1467 & 18.6 & 0.5 & 0.55 & 0.01 & 2.261 & 0.130 & 535.4 & 1.2 & -0.072 & 0.034 & -1.63 & 0.19 & 458.8 & 4.5 & 10.48 & 1.15 & 3.39 & 0.11 & 1.298 & 1243 \\
BATSE 1580 & 16.2 & 0.7 & 0.90 & 0.06 & 1.209 & 0.115 & 526.0 & 1.7 & 0.100 & 0.054 & -0.43 & 0.03 & 210.3 & 5.5 & 0.14 & 0.04 & 4.89 & 0.23 & 1.226 & 931 \\
BATSE 1806 & 42.8 & 5.2 & 0.78 & 0.04 & 0.516 & 0.109 & 501.4 & 1.0 & 0.062 & 0.029 & -4.74 & 0.52 & 62.8 & 2.6 & 2.52 & 1.11 & 11.16 & 1.49 & 1.192 & 1556 \\
BATSE 1883 & 8.8 & 0.1 & 0.70 & 0.01 & 5.200 & 0.184 & 537.3 & 0.6 & -0.039 & 0.008 & -0.51 & 0.03 & 1014.5 & 8.0 & 1.31 & 0.09 & 2.05 & 0.04 & 1.183 & 3118 \\
BATSE 2662 & 19.3 & 0.6 & 0.86 & 0.04 & 1.512 & 0.143 & 579.6 & 0.4 & 0.090 & 0.006 & -0.52 & 0.05 & 245.5 & 4.6 & 0.40 & 0.08 & 5.52 & 0.20 & 1.087 & 3899 \\
BATSE 2665 & 20.7 & 0.7 & 0.84 & 0.03 & 1.986 & 0.154 & 681.3 & 0.5 & -0.086 & 0.002 & -0.74 & 0.06 & 285.5 & 5.2 & 0.53 & 0.10 & 5.82 & 0.21 & 1.308 & 8180 \\
BATSE 2862 & 4.0 & 0.4 & 0.88 & 0.04 & 0.914 & 0.116 & 506.0 & 0.9 & -0.173 & 0.010 & -0.29 & 0.03 & 151.1 & 9.4 & 0.06 & 0.04 & 1.16 & 0.14 & 1.064 & 2493 \\
BATSE 2958p1 & 14.1 & 0.7 & 0.59 & 0.02 & 3.751 & 0.189 & 566.0 & 0.5 & -0.133 & 0.004 & -32.86 & 0.23 & 240.2 & 4.4 & 5.48 & 1.13 & 2.78 & 0.17 & 1.119 & 4676 \\
BATSE 3003 & 30.8 & 0.7 & 0.49 & 0.01 & 2.826 & 0.159 & 485.1 & 3.6 & 0.005 & 0.009 & -2.16 & 0.33 & 485.1 & 3.6 & 28.63 & 2.57 & 5.02 & 0.14 & 1.412 & 3118 \\
BATSE 3026 & 11.2 & 1.0 & 0.56 & 0.03 & 1.044 & 0.121 & 652.2 & 0.7 & -0.036 & 0.008 & -2.48 & 0.36 & 166.9 & 4.8 & 5.73 & 1.99 & 2.08 & 0.22 & 1.038 & 3118 \\
BATSE 3040 & 19.6 & 0.6 & 0.91 & 0.04 & 1.699 & 0.128 & 578.0 & 0.8 & -0.144 & 0.015 & -0.38 & 0.03 & 277.6 & 5.2 & 0.14 & 0.03 & 5.93 & 0.19 & 1.177 & 2102 \\
BATSE 3143* & 5.6 & 0.2 & 0.67 & 0.03 & 2.589 & 0.144 & 622.5 & 0.4 & -0.003 & 0.006 & -0.56 & 0.05 & 475.1 & 8.2 & 1.04 & 0.18 & 1.26 & 0.06 & 1.090 & 3899 \\
BATSE 3168 & 27.1 & 6.5 & 0.26 & 0.01 & 0.507 & 0.103 & 551.9 & 0.4 & 0.019 & 0.005 & -23.54 & 7.23 & 71.3 & 2.5 & 248.34 & 226.72 & 2.36 & 0.71 & 1.029 & 3899 \\
BATSE 3257 & 50.4 & 0.7 & 0.88 & 0.01 & 3.063 & 0.133 & 574.8 & 0.5 & 0.062 & 0.006 & -0.35 & 0.04 & 409.5 & 3.3 & 0.65 & 0.05 & 14.86 & 0.21 & 1.138 & 3899 \\
BATSE 7567* & 3.3 & 0.2 & 0.68 & 0.05 & 2.372 & 0.112 & 695.9 & 0.7 & 0.005 & 0.014 & -0.41 & 0.04 & 514.7 & 12.1 & 0.58 & 0.13 & 0.75 & 0.05 & 1.251 & 2337 \\
BATSE 7614 & 31.4 & 1.4 & 0.66 & 0.01 & 1.224 & 0.104 & 978.1 & 0.9 & -0.033 & 0.017 & -3.40 & 0.32 & 224.2 & 3.6 & 6.66 & 1.15 & 6.88 & 0.34 & 1.007 & 2206 \\
BATSE 7638 & 18.5 & 0.5 & 0.85 & 0.09 & 1.750 & 0.109 & 620.8 & 0.7 & 0.112 & 0.013 & -0.50 & 0.04 & 318.9 & 5.0 & 0.43 & 0.07 & 5.26 & 0.16 & 1.134 & 2337 \\
BATSE 7711 & 14.4 & 0.3 & 0.71 & 0.01 & 3.663 & 0.127 & 678.2 & 0.7 & 0.144 & 0.013 & -0.83 & 0.05 & 650.6 & 5.9 & 1.85 & 0.02 & 3.41 & 0.07 & 1.116 & 2337 \\
BATSE 7775* & 1.0 & 0.1 & 0.77 & 0.17 & 3.400 & 0.140 & 581.2 & 0.4 & -0.065 & 0.004 & -0.10 & 0.01 & 696.7 & 23.1 & 0.07 & 0.02 & 0.26 & 0.02 & 1.041 & 4680 \\
BATSE 7843 & 15.5 & 0.8 & 0.73 & 0.03 & 1.354 & 0.100 & 636.7 & 0.5 & -0.061 & 0.006 & -1.48 & 0.14 & 199.2 & 4.6 & 1.65 & 0.37 & 3.75 & 0.22 & 1.074 & 3899 \\
BATSE 7903 & 4.6 & 0.3 & 0.73 & 0.06 & 1.814 & 0.107 & 636.2 & 0.4 & -0.021 & 0.006 & -0.44 & 0.05 & 328.9 & 8.8 & 0.47 & 0.12 & 1.12 & 0.07 & 1.056 & 3899 \\
BATSE 7989 & 11.3 & 0.3 & 0.95 & 0.01 & 2.988 & 0.134 & 500.8 & 0.4 & -0.002 & 0.005 & -0.27 & 0.01 & 325.4 & 7.6 & 0.02 & 0.00 & 3.59 & 0.11 & 1.113 & 3899 \\
BATSE 8112 & 10.4 & 2.6 & 0.31 & 0.03 & 0.682 & 0.090 & 555.9 & 0.6 & -0.175 & 0.008 & -7.70 & 2.36 & 95.1 & 4.3 & 52.00 & 50.73 & 1.08 & 0.33 & 1.063 & 3118 \\
BATSE 8121 & 48.2 & 1.4 & 0.68 & 0.01 & 1.167 & 0.095 & 862.6 & 0.9 & -0.125 & 0.010 & -7.48 & 0.29 & 269.1 & 3.0 & 7.95 & 0.90 & 11.00 & 0.35 & 1.099 & 3118 \\
GRB081224(GBM)$\dagger$ & 15.6 & 0.6 & 0.67 & 0.02 & 2.14 & 0.55 & 55.2 & 0.3 & 0.019 & 0.015 & -0.90 & 0.12 & 138.3 & 2.4 & 2.86 & 0.43 & 3.50 & 0.14 & 1.484 & 1024 \\
GRB100707a(GBM)$\dagger$ & 19.2 & 0.3 & 0.83 & 0.02 & 31.24 & 0.84 & 51.9 & 0.4 & 0.181 & 0.017 & -0.15 & 0.03 & 278.7 & 3.1 & 0.64 & 0.06 & 5.29 & 0.10 & 1.564 & 999 \\
GRB120326a(GBM)$\dagger$ & 13.7 & 1.8 & 0.54 & 0.03 & 8.62 & 0.59 & 52.2 & 0.3 & -0.004 & 0.013 & -3.09 & 0.68 & 38.5 & 1.7 & 8.22 & 4.15 & 2.47 & 0.38 & 1.457 & 1076 \\
\tableline
BATSE 0469 & 8.1 & 0.2 & 0.80 & 0.02 & 4.222 & 0.198 & 793.5 & 0.7 & 0.105 & 0.007 & 1.61 & 0.02 & 965.1 & 12.2 & 0.39 & 0.04 & 2.17 & 0.05 & 1.894 & 4680 \\
BATSE 0795 & 26.9 & 1.2 & 0.76 & 0.02 & 1.115 & 0.129 & 484.6 & 0.6 & -0.052 & 0.008 & -1.76 & 0.17 & 169.8 & 3.3 & 2.1 & 0.4 & 6.78 & 0.32 & 1.19 & 3118 \\
BATSE 2958p2 & 10.3 & 0.4 & 0.44 & 0.01 & 3.751 & 0.189 & 566.0 & 0.5 & -0.133 & 0.004 & -4.07 & 0.22 & 535.2 & 5.7 & 15.27 & 2.20 & 1.50 & 0.07 & 1.119 & 4676 \\
GRB130427a(GBM)p1$\dagger$ & 4.0 & 0.2 & 0.77 & 0.02 & p256 & sp256 & 476.4 & 16.5 & 7.6 & 5.6 & -0.12 & 0.01 & 3381.1 & 61.4 & 0.26 & 0.04 & 1.04 & 0.05 & 10.98 & 112 \\
GRB130427a(GBM)p2$\dagger$ & 98.1 & 2.0 & 0.80 & 0.01 & p256 & sp256 & 560.5 & 2.3 & -0.3 & 0.1 & 118.6 & 0.2 & 218.2 & 2.1 & 4.81 & 0.41 & 26.06 & 0.58 & 1.135 & 3118 \\
\enddata

\tablecomments{Column 1 is the Burst ID (Intermediate GRBs are denoted by asterisks; all others are Long. Classifications are obtained from \cite{hor06} when available, and are otherwise estimated from pulse properties using the classification criteria of \cite{hak03}; denoted by $\dagger$), column 2 the pulse base duration $w$ (s), column 3 the base duration uncertainty $\sigma_w$ (s), column 4 the base asymmetry $\kappa$, column 5 the base asymmetry uncertainty columns $\sigma_\kappa$  column 6 the burst's 256 ms peak flux $p_{\rm 256}$ ($cm^{-2} s^{-1}$), column 7 the 256 ms peak flux uncertainty $\sigma_{p_{\rm 256}}$ ($cm^{-2} s^{-1}$), column 7 the fitted background counts per 64 ms bin $B$, column 8 the background uncertainty $\sigma_B$, column 9 the rate of change in the background counts per 64 ms bin $S_B$ ($s^{-1}$), column 10 the rate of change in the background counts per 64 ms bin uncertainty $\sigma_{S_B}$ ($s^{-1}$), column 11 the pulse start time $t_s$ (s), column 12 the pulse start time uncertainty $\sigma_{t_s}$ (s), column 13 the pulse amplitude $A$, column 14 the pulse amplitude uncertainty $\sigma_A$ (s), column 15 the pulse rise parameter $\tau_1$, column 16 the pulse rise parameter uncertainty $\sigma_{\tau_1}$, column 17 the pulse decay parameter $\tau_2$, column 18 the pulse decay parameter uncertainty $\sigma_{\tau_2}$, column 18 the $\chi^2_\nu$ per degree of freedom $\nu$. for the entire fit (all pulses plus background in selected interval), and column 19 the number of degrees of freedom $\nu$.}
\end{deluxetable}

\clearpage

\begin{table}
\scriptsize
\begin{center}
\caption{Fit parameters for GRB pulse template residuals.\label{tbl-2}}
\begin{tabular}{crrrrrrrrrrrrr}
\tableline\tableline
Template & $\langle \kappa \rangle$ & $N_{\rm pulses}$ &  $t_0$ & $\sigma_{t_0}$ & $A$ & $\sigma_A$ & $\Omega$ & $\sigma_\Omega$ & $s$ & $\sigma_s$  & $\chi^2_{\nu}$ \\
\tableline
s & 0.27 & 12 & 0.472 & 0.001 & 0.0245 & 0.0005 & 200 &  4 & 0.75 & 0.02 & 1.0 \\
a & 0.58 & 10 & 0.212 & 0.001 & 0.0408 & 0.0004& 400 & 10 & 0.29 & 0.01 & 1.8 \\
v & 0.73 & 13 & 0.154 & 0.001 & 0.0455 & 0.0004 & 748 & 17 & 0.16 & 0.01 & 1.7 \\
x & 0.87 & 12 & 0.116 & 0.001 & 0.0545 & 0.0004 &1623 & 42& 0.10 & 0.01 & 1.2 \\
\tableline
\end{tabular}
\end{center}

\normalsize
\end{table}

\clearpage

\begin{table}

\scriptsize

\begin{center}
\caption{GRB pulse residual fit parameters.\label{tbl-3}}
\begin{tabular}{crrrrrrrrrrrr}
\tableline
\tableline
GRB pulse & $t_0$ & $\sigma_{t_0}$ & $A$ & $\sigma_A$ & $\Omega$ & $\sigma_\Omega$ & $s$ & $\sigma_s$  \\
\tableline
BATSEÊ 332 &  3.21 &   0.09 &  36.4 &   1.1 &  12.47 &   0.97 &  0.095 &  0.008 \\
BATSEÊ 493 & 0.90 &   0.04 &  48.4 &   4.5 &  22.09 &   1.93 &  0.896 &  0.099 \\
BATSEÊ 501 & 0.02 &   0.01 &  43.8 &   3.4 &  57.87 &   9.04 &  0.187 &  0.031 \\
BATSEÊ 540 &-0.19 &   0.02 &  93.0 &   6.0 &  78.30 &  12.16 &  0.177 &  0.030  \\
BATSEÊ 563 & 1.34 &   0.05 &  55.4 &   1.9 &  21.48 &   2.04 &  0.074 &  0.007 \\
BATSEÊ 658 & 0.73 &   0.03 &  44.2 &   2.4 &  14.17 &   1.00 &  0.686 &  0.063 \\
BATSE  673 &  5.55 &   0.15 &  15.0 &   1.4 &  10.50 &   1.31 &  0.390 &  0.057 \\
BATSE  680 & 1.00 &   0.06 &  36.6 &   2.5 &  18.20 &   1.60 &  0.361 &  0.040 \\
BATSE  711 &-0.10 &   0.03 &  56.2 &   5.5 &  83.99 &  17.68 &  0.475 &  0.111 \\
BATSE  727 & 0.68 &   0.06 &  40.1 &   4.2 &  19.49 &   1.87 &  0.577 &  0.073 \\
BATSEÊ 795 & 2.29 &   0.06 &  52.0 &   1.6 &  11.84 &   0.69 &  0.235 &  0.015 \\
BATSEÊ 907 & 1.10 &   0.03 &  71.4 &   2.7 &  32.04 &   3.42 &  0.106 &  0.012 \\
BATSEÊ 914 &  0.48 &   0.03 &  51.7 &   3.3 &  42.30 &   5.50 &  0.231 &  0.032 \\
BATSE 1039 & 4.10 &   0.05 &  55.3 &   2.0 &  10.56 &   0.51 &  0.326 &  0.019 \\
BATSE 1145 & 0.47 &   0.02 &  83.9 &   8.7 &  38.15 &   2.79 &  0.655 &  0.064 \\
BATSE 1200 & 5.62 &   0.10 &  48.0 &   1.6 &   4.08 &   0.16 &  0.811 &  0.044 \\
BATSE 1301 & 0.67 &   0.21 &   7.9 &   1.3 &   5.13 &   0.87 &  0.611 &  0.135 \\
BATSE 1306 &-0.72 &   0.04 &  44.2 &   3.4 &  22.86 &   2.14 &  0.686 &  0.086 \\
BATSE 1319 &-0.68 &   0.02 &  97.4 &   4.4 &  27.83 &   1.72 &  0.459 &  0.033 \\
BATSE 1379 &-0.14 &   0.02 &  83.6 &   7.3 &  62.15 &   6.90 &  0.861 &  0.120 \\
BATSE 1406 & 2.73 &   0.09 &  35.2 &   1.3 &  13.44 &   1.17 &  0.097 &  0.009 \\
BATSE 1432 & 1.60 &   0.14 &  25.3 &   1.5 &   2.86 &   0.17 &  0.849 &  0.071 \\
BATSE 1446 & 1.50 &   0.12 &  15.6 &   0.7 &   4.49 &   0.35 &  0.420 &  0.042 \\
BATSE 1467 & 4.17 &   0.05 &  76.1 &   1.9 &  10.72 &   0.45 &  0.310 &  0.015 \\
BATSE 1580 & 1.07 &   0.06 &  30.4 &   2.0 &  30.61 &   4.87 &  0.108 &  0.018 \\
BATSE 1806 & 1.23 &   0.10 &  24.7 &   1.6 &  10.22 &   1.25 &  0.219 &  0.029 \\
BATSE 1883 & 1.13 &   0.03 &  74.1 &   3.4 &  35.13 &   3.24 &  0.148 &  0.014 \\
BATSE 2662 & 1.25 &   0.07 &  31.0 &   1.4 &  21.62 &   3.01 &  0.152 &  0.023 \\
BATSE 2665 & 1.53 &   0.04 &  85.0 &   2.5 &  20.10 &   1.19 &  0.163 &  0.010 \\
BATSE 2862 & 0.01 &   0.02 &  46.3 &   2.8 & 144.82 &  34.45 &  0.114 &  0.029 \\
BATSE 2958p1 &-29.35 &   0.04 &  30.3 &   0.6 &  20.58 &   1.34 &  0.238 &  0.017 \\
BATSE 3003 & 8.72 &   0.06 &  91.8 &   1.5 &   5.79 &   0.15 &  0.340 &  0.010 \\
BATSE 3026 & 1.31 &   0.04 &  61.8 &   2.8 &  18.01 &   1.30 &  0.408 &  0.035 \\
BATSE 3040 & 0.85 &   0.02 &  77.9 &   2.6 &  40.16 &   3.24 &  0.085 &  0.007 \\
BATSE 3143 & 0.63 &   0.03 &  70.7 &   5.3 &  39.06 &   3.42 &  0.307 &  0.031 \\
BATSE 3168 & 1.31 &   0.10 &  22.5 &   1.1 &   3.50 &   0.19 &  0.861 &  0.063 \\
BATSE 3257 & 2.30 &   0.06 &  48.7 &   1.2 &  19.07 &   1.53 &  0.055 &  0.005 \\
BATSE 7567 & 0.26 &   0.02 & 142.8 &   7.9 &  72.65 &  11.65 &  0.160 &  0.027 \\
BATSE 7614 & 2.68 &   0.08 &  39.8 &   3.2 &  16.21 &   1.72 &  0.342 &  0.043 \\
BATSE 7638 & 1.06 &   0.03 &  58.4 &   2.5 &  32.78 &   3.18 &  0.171 &  0.018 \\
BATSE 7711 & 2.52 &   0.08 &  46.1 &   2.1 &  16.57 &   1.65 &  0.221 &  0.024 \\
BATSE 7775 & 0.05 &   0.02 &  56.3 &  10.1 & 363.11 & 222.19 &  0.155 &  0.098 \\
BATSE 7843 & 1.21 &   0.05 &  50.9 &   2.4 &  17.02 &   1.35 &  0.303 &  0.027 \\
BATSE 7903 & 0.24 &   0.02 &  86.7 &   6.3 &  86.31 &   9.03 &  0.192 &  0.024 \\
BATSE 7989 & 0.11 &   0.02 &  67.8 &   3.4 & 156.32 &  23.80 &  0.043 &  0.007 \\
BATSE 8112 &-0.47 &   0.04 &  28.0 &   2.1 &  10.08 &   0.79 &  0.969 &  0.107 \\
BATSE 8121 & 2.55 &   0.09 &  46.2 &   1.7 &   7.02 &   0.42 &  0.254 &  0.017 \\
GRB081224(GBM)  & 2.76 &   0.08 &  17.5 &   1.0 &  15.69 &   1.39 &  0.262 &  0.027 \\
GRB100707a(GBM) & 1.66 &   0.02 &  52.1 &   1.1 &  27.04 &   1.20 &  0.133 &  0.006 \\
GRB120326a(GBM) & 1.24 &   0.08 &   8.7 &   1.0 &  15.00 &   2.08 &  0.684 &  0.117 \\
\tableline
BATSE 0469 & 2.57 &   0.03 & 309.4 &  11.2 &  30.19 &   1.58 &  0.287 &  0.017 \\
BATSE 3164 &-0.18 &   0.03 &  43.8 &   4.1 &  64.93 &  10.99 &  0.114 &  0.020 \\
BATSE 2958p2 & 0.09 &   0.01 & 113.5 &   5.2 &  25.45 &   93.3 &  510 &  0.18 \\
GRB130427a(GBM)p1 & 0.38 &   0.02 &  628.1 &   39.0 & 100.00 &  14.99 &  0.237 &  0.039 \\
GRB130427a(GBM)p2 & 127.51 &   0.09 &   37.7 &    2.0 &  12.13 &   1.00 &  0.348 &  0.034 \\
\tableline
\end{tabular}
\end{center}

\normalsize
\end{table}

\clearpage

\begin{table}
\scriptsize
\begin{center}
\caption{Fit parameters for GRB pulse residuals.\label{tbl-4}}
\begin{tabular}{crrrrrrrr}
\tableline\tableline
GRB pulse & $t_{\rm start}$ & $t_{\rm end}$ & $\chi^2_{\rm p}$ & $\nu_{\rm p}$ & $\chi^2_{\nu; \rm p}$ & $\chi^2_{\rm p+r}$ & $\nu_{\rm p+r}$ & $\chi^2_{\nu; \rm p+r}$\\
\tableline
BATSEÊ 332 & -9.175 &   81.527 & 1700.9 & 1411 &  1.21 & 1471.8 & 1407 &  1.05 \\
BATSEÊ 493 & -2.485 &    6.370 &  161.7 &  132 &  1.23 &  124.2 &  128 &  0.97 \\
BATSEÊ 501 & -3.020 &   21.543 &  472.3 &  378 &  1.25 &  438.3 &  374 &  1.17 \\
BATSEÊ 540 & -1.581 &    8.511 &  256.4 &  151 &  1.70 &  194.4 &  147 &  1.32 \\
BATSEÊ 563 & -7.816 &   71.947 & 1633.7 & 1241 &  1.32 & 1478.4 & 1237 &  1.20 \\
BATSEÊ 658 & -7.605 &   34.809 &  773.6 &  656 &  1.18 &  713.5 &  652 &  1.09 \\
BATSE  673 &-55.725 &   47.093 &  741.3 & 1600 &  0.46 &  721.5 & 1596 &  0.45\\
BATSE  680 &-11.227 &   10.411 &  258.3 &  332 &  0.78 &  218.7 &  328 &  0.67 \\
BATSE  711 & -1.729 &    2.326 &   72.4 &   58 &  1.25 &  57.2 &   54 &  1.06 \\
BATSE  727 & -4.552 &    7.783 &  147.1 &  187 &  0.79 &  145.0 &  183 &  0.79 \\
BATSEÊ 795 & -7.337 &   57.797 & 1382.0 & 1012 &  1.37 & 1192.0 & 1008 &  1.18 \\
BATSEÊ 907 & -3.891 &   35.972 & 1087.2 &  617 &  1.76 &  889.2 &  613 &  1.45 \\
BATSEÊ 914 & -1.989 &   11.822 &  248.1 &  210 &  1.18 &  199.4 &  206 &  0.97 \\
BATSE 1039 & -8.046 &   27.376 &  607.7 &  547 &  1.11 &  474.2 &  543 &  0.87 \\
BATSE 1145 & -1.608 &    5.434 &  215.4 &  104 &  2.07 &  149.7 &  100 &  1.50 \\
BATSE 1200 &-13.563 &   31.068 &  907.7 &  691 &  1.31 &  762.1 &  687 &  1.11 \\
BATSE 1301 &-34.448 &   28.239 &  528.3 &  904 &  0.58 &  522.0 &  900 &  0.58 \\
BATSE 1306 & -9.124 &   13.851 &  286.4 &  353 &  0.81 &  267.7 &  349 &  0.77 \\
BATSE 1319 & -4.470 &   16.763 &  443.4 &  325 &  1.36 &  337.8 &  321 &  1.05 \\
BATSE 1379 & -1.522 &    4.994 &  162.5 &   96 &  1.69 &  127.3 &   92 &  1.38 \\
BATSE 1406 & -7.390 &   62.050 & 1328.2 & 1079 &  1.23 & 1169.0 & 1075 &  1.09 \\
BATSE 1432 &-26.101 &  101.185 & 1790.5 & 1983 &  0.90 & 1730.6 & 1979 &  0.87 \\
BATSE 1446 &-65.348 &   29.644 &  674.6 & 1479 &  0.46 &  635.1 & 1475 &  0.43 \\
BATSE 1467 & -4.418 &   32.180 &  844.0 &  566 &  1.49 &  597.3 &  562 &  1.06 \\
BATSE 1580 & -4.376 &   39.819 &  787.9 &  685 &  1.15 &  725.3 &  681 &  1.07 \\
BATSE 1806 &-14.112 &   94.329 & 1859.7 & 1688 &  1.10 & 1789.0 & 1684 &  1.06 \\
BATSE 1883 & -2.184 &   17.914 &  561.7 &  308 &  1.82 &  417.8 &  304 &  1.37 \\
BATSE 2662 & -4.986 &   45.586 &  852.4 &  784 &  1.09 &  804.2 &  780 &  1.03 \\
BATSE 2665 & -5.470 &   48.287 & 1098.0 &  834 &  1.32 &  957.5 &  830 &  1.15 \\
BATSE 2862 & -1.252 &    9.555 &  212.7 &  163 &  1.31 &  187.0 &  159 &  1.18 \\
BATSE 2958p1& -38.327 &   25.619 &  136.6 &  514 &  0.27 &   94.1 &  508 &  0.19 \\
BATSE 3003 & -6.157 &   49.825 & 1819.6 &  869 &  2.09 & 1080.7 &  865 &  1.25 \\
BATSE 3026 & -4.340 &   19.556 &  461.5 &  367 &  1.26 &  367.7 &  363 &  1.01 \\
BATSE 3040 & -5.163 &   48.380 & 1177.1 &  830 &  1.42 &  867.2 &  826 &  1.05 \\
BATSE 3143 & -1.620 &   11.139 &  284.1 &  194 &  1.46 &  231.1 &  190 &  1.22 \\
BATSE 3168 &-26.761 &   32.815 &  597.1 &  925 &  0.65 &  570.8 &  921 &  0.62 \\
BATSE 3257 &-12.263 &  121.886 & 2374.6 & 2090 &  1.14 & 2082.0 & 2086 &  1.00 \\
BATSE 7567 & -1.040 &    6.582 &  559.6 &  113 &  4.95 &  387.5 &  109 &  3.56 \\
BATSE 7614 & -9.180 &   61.171 & 1077.1 & 1093 &  0.99 & 1037.1 & 1089 &  0.95 \\
BATSE 7638 & -4.758 &   43.543 &  891.0 &  749 &  1.19 &  823.6 &  745 &  1.11 \\
BATSE 7711 & -3.622 &   29.591 &  673.4 &  513 &  1.31 &  546.2 &  509 &  1.07 \\
BATSE 7775 & -0.313 &    2.200 &   44.3 &   33 &  1.34 &   33.8 &   29 &  1.17 \\
BATSE 7843 & -4.611 &   32.327 &  623.0 &  572 &  1.09 &  567.4 &  568 &  1.00 \\
BATSE 7903 & -1.372 &    9.631 &  231.6 &  166 &  1.39 &  166.2 &  162 &  1.03 \\
BATSE 7989 & -3.160 &   28.953 &  749.6 &  496 &  1.51 &  611.6 &  492 &  1.24 \\
BATSE 8112 & -9.059 &   13.522 &  251.7 &  347 &  0.73 &  229.4 &  343 &  0.67 \\
BATSE 8121 &-16.944 &   96.559 & 1809.2 & 1767 &  1.02 & 1699.5 & 1763 &  0.96 \\
GRB081224(GBM)  & -3.758 &   30.854 &  662.4 &  535 &  1.24 &  436.6 &  362 &  1.21 \\
GRB100707a(GBM) & -4.387 &   44.050 & 1316.5 &  751 &  1.75 &  974.3 &  747 &  1.30 \\
GRB120326a(GBM) & -5.294 &   23.503 &  351.9 &  390 &  0.90 &  350.6 &  386 &  0.91 \\
\tableline
BATSE 0469 & 0.040 &   18.247 & 3466.6 &  279 & 12.43 & 2259.2 &  275 &  8.22 \\
BATSE 3164 &-2.359 &   14.837 &  353.4 &  262 &  1.35 &  311.5 &  258 &  1.21 \\
BATSE 2958p2 & -5.567 &   15.696 &  536.2 &  326 &  1.64 &  347.2 &  322 &  1.08 \\
GRB130427a(GBM)p1 & -0.964 &    8.841 & 1166.3 & 50 & 23.33 &  742.8 & 46 & 16.15 \\
GRB130427a(GBM)p2 &  109.693 &  219.163 & 1913.1 & 1705 &  1.12 & 1856.1 & 1701 &  1.09 \\
\tableline
\end{tabular}
\end{center}

\normalsize
\end{table}

\clearpage

\begin{figure}
  \includegraphics{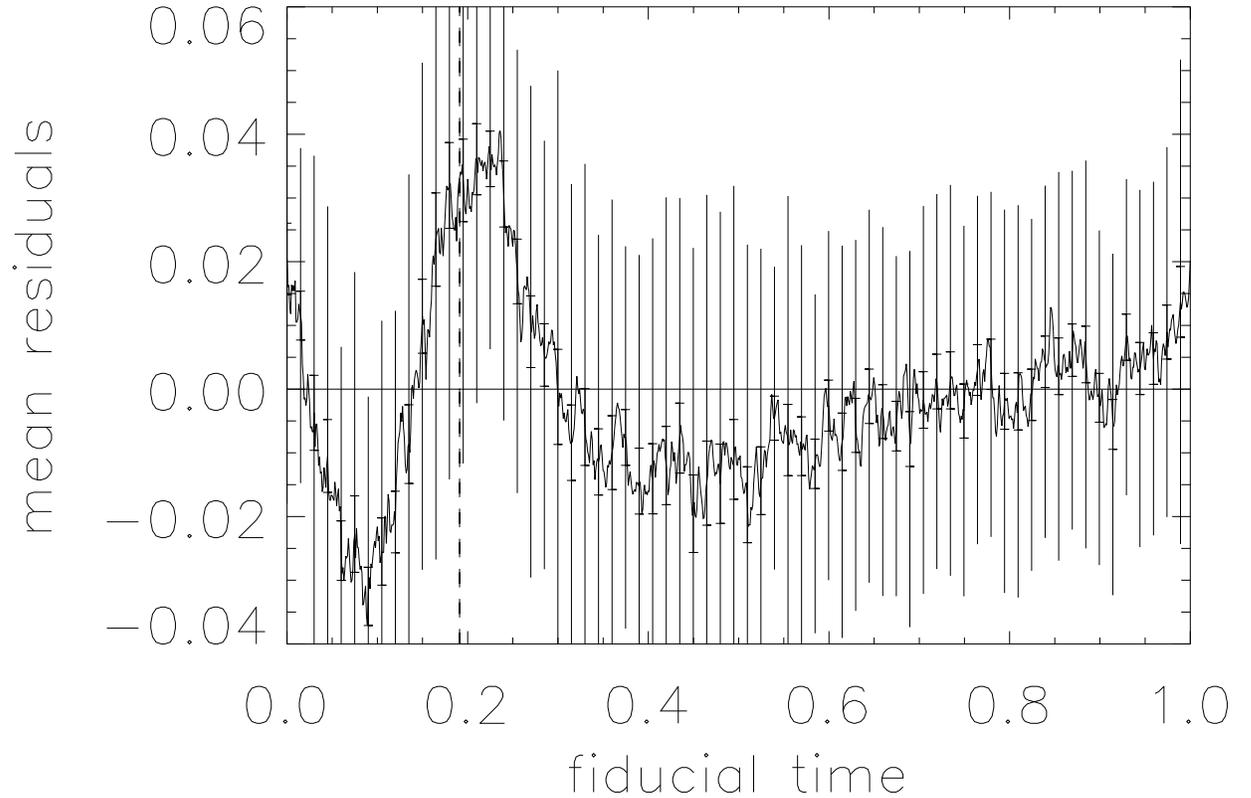}
  \caption{Mean pulse fit residuals for summed 20 keV - 1 MeV data for 48 of the 50 GRB pulses described in the text and scaled to a fiducial timescale. The time interval shown corresponds to the base duration interval, occurring between times when each pulse intensity is $A_i \exp^{-3}$ ($A_i$ is a pulse'a peak intensity). The mean residuals systematically differ from zero, indicating that a generic pulse shape exists that differs from the Norris et al.\ (2005) pulse shape. The fitted pulse peak is found at normalized phase (fiducial time) $\langle \Phi_{pk} \rangle=0.191$ (dashed line).\label{fig1}}
\end{figure}

\clearpage

\begin{figure}
  \includegraphics{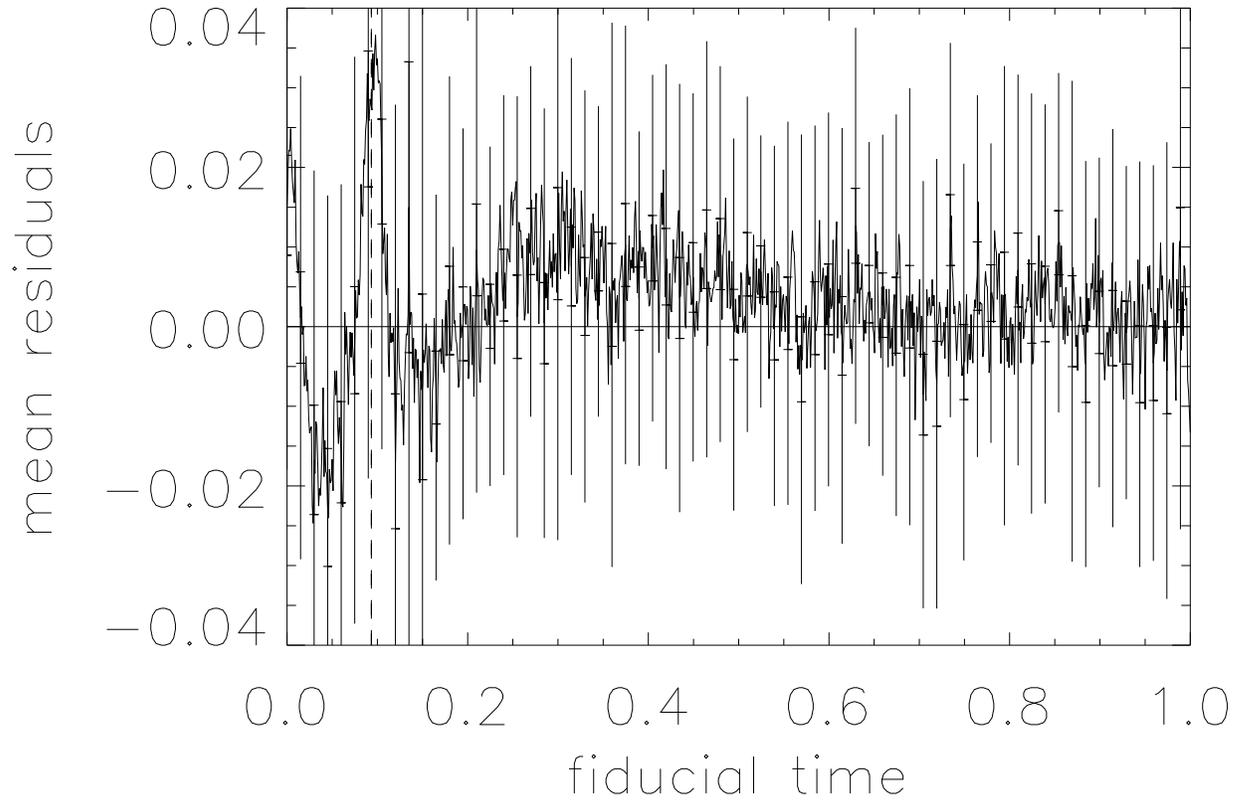}
  \caption{Mean pulse fit residuals for summed 20 keV - 1 MeV data for a subsample of 28 pulses and a fiducial timescale defined when the pulse intensity is $A \exp^{-20}$. The fitted pulse peak is found at fiducial time $\langle \Phi_{pk} \rangle=0.093$ (dashed line). The excess occurring on the pulse decay has declined to zero well before the end of the sampling time is reached.\label{fig2}}
\end{figure}

\clearpage

\begin{figure}
  \includegraphics{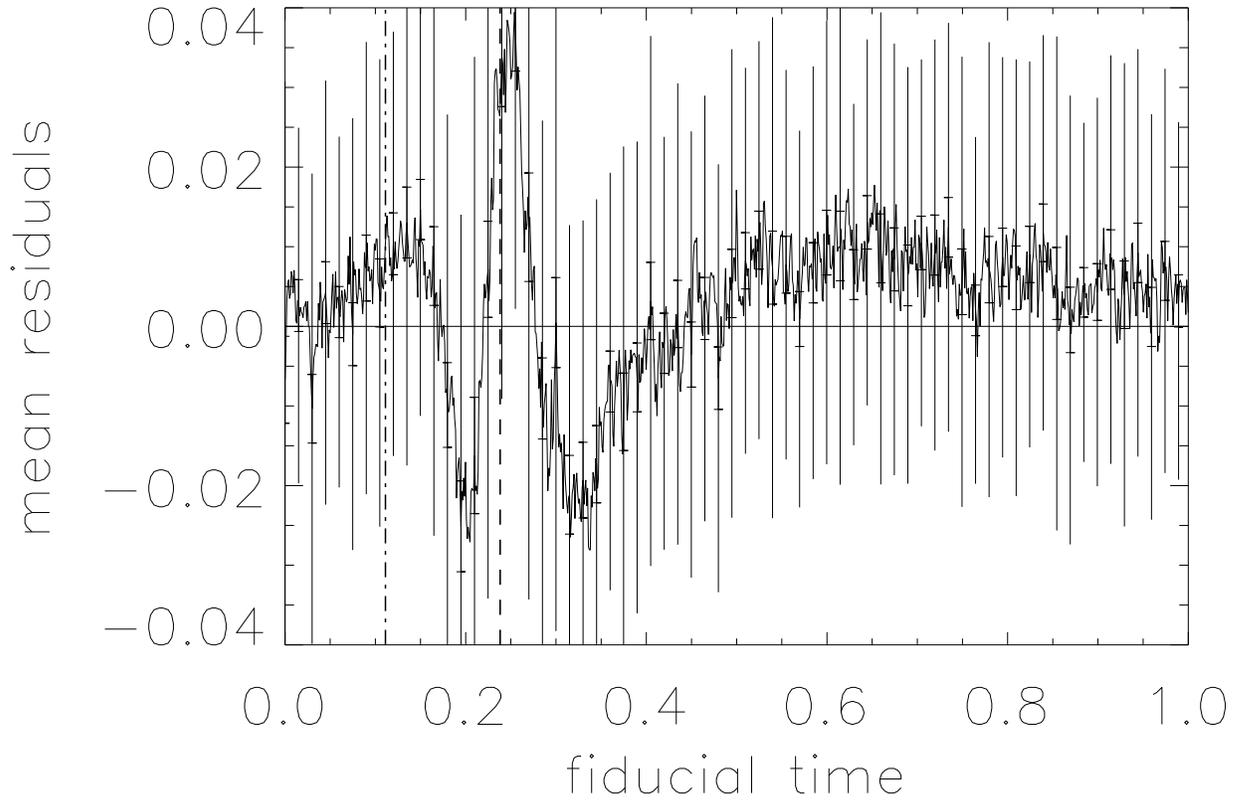}
  \caption{Pulse fit residuals for summed 20 keV - 1 MeV data for the 48 pulse sample; the fiducial timescale is defined when the decay of the fitted pulse has reached $A \exp^{-8}$, while the rise is $0.1 \times$ this decay duration. The mean fitted pulse peak is found at fiducial time $\langle \Phi_{pk} \rangle=0.237$ (dashed line). The mean residual distribution undergoes three brightening periods.\label{fig3}}
\end{figure}

\clearpage

\begin{figure}
  \includegraphics{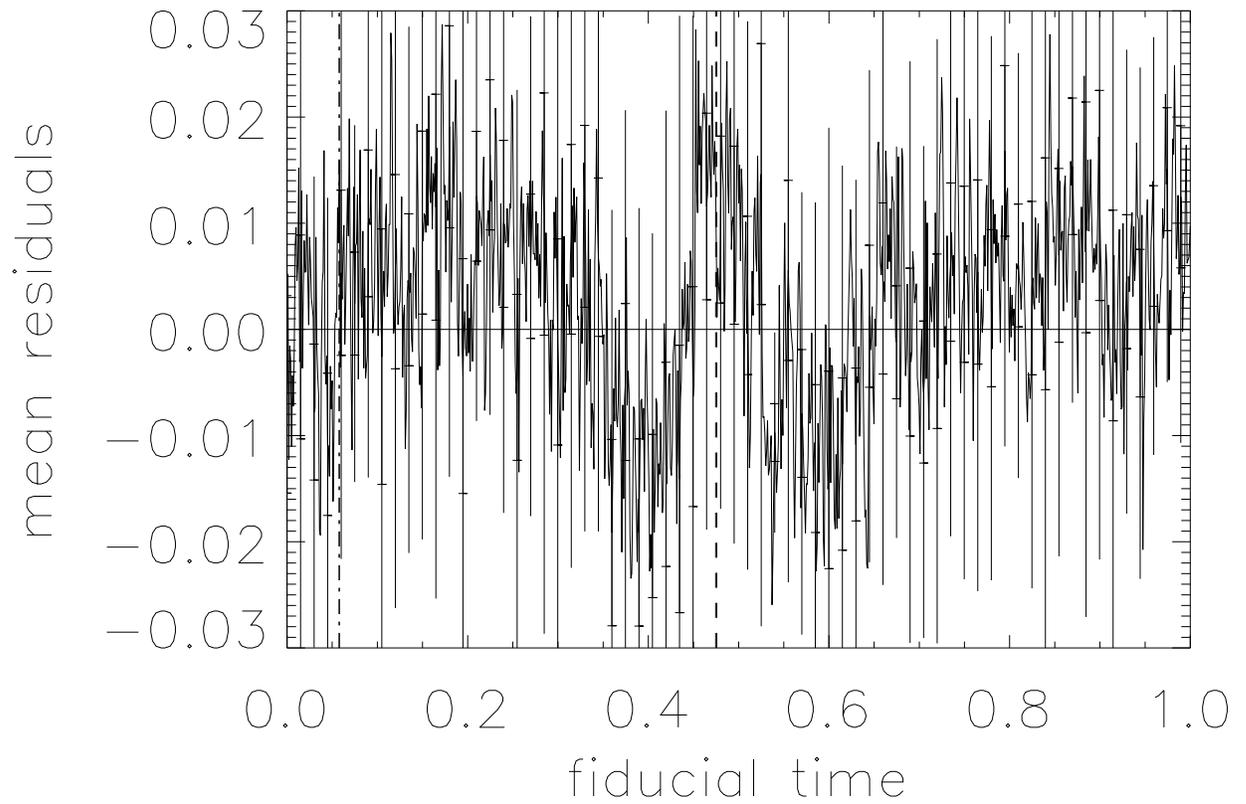}
  \caption{Pulse fit residuals for 12 symmetric (`s') pluses having $\kappa < 0.45$. The mean fitted pulse peak is found at fiducial time $\langle \Phi_{pk} \rangle=0.476$ (dashed line). \label{fig4}}
\end{figure}

\clearpage

\begin{figure}
  \includegraphics{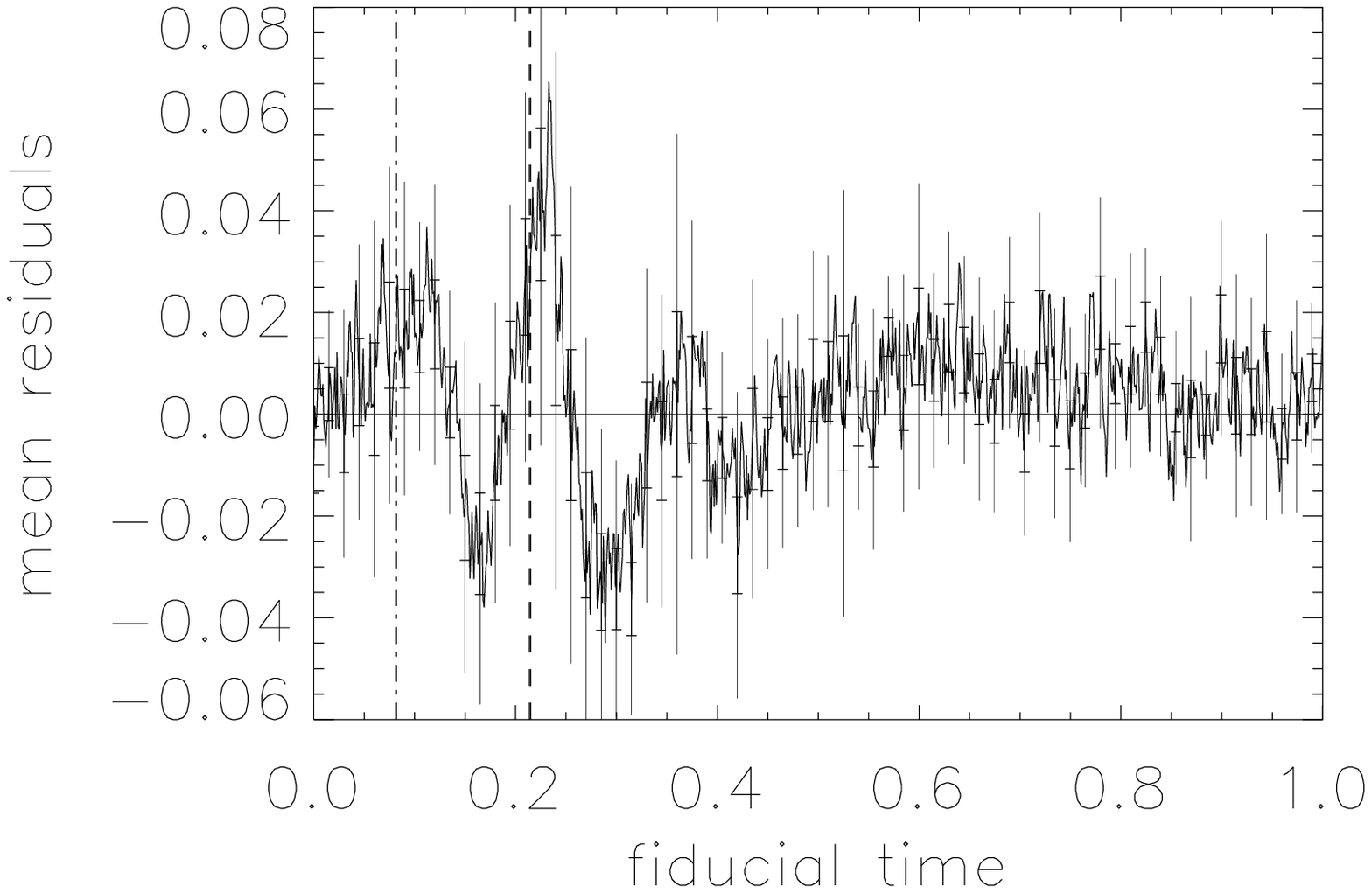}
  \caption{Pulse fit residuals for 10 asymmetric (`a') pluses having $0.45 \le \kappa < 0.67$. The mean fitted pulse peak is found at fiducial time $\langle \Phi_{pk} \rangle=0.215$ (dashed line). \label{fig5}}
\end{figure}

\clearpage

\begin{figure}
  \includegraphics{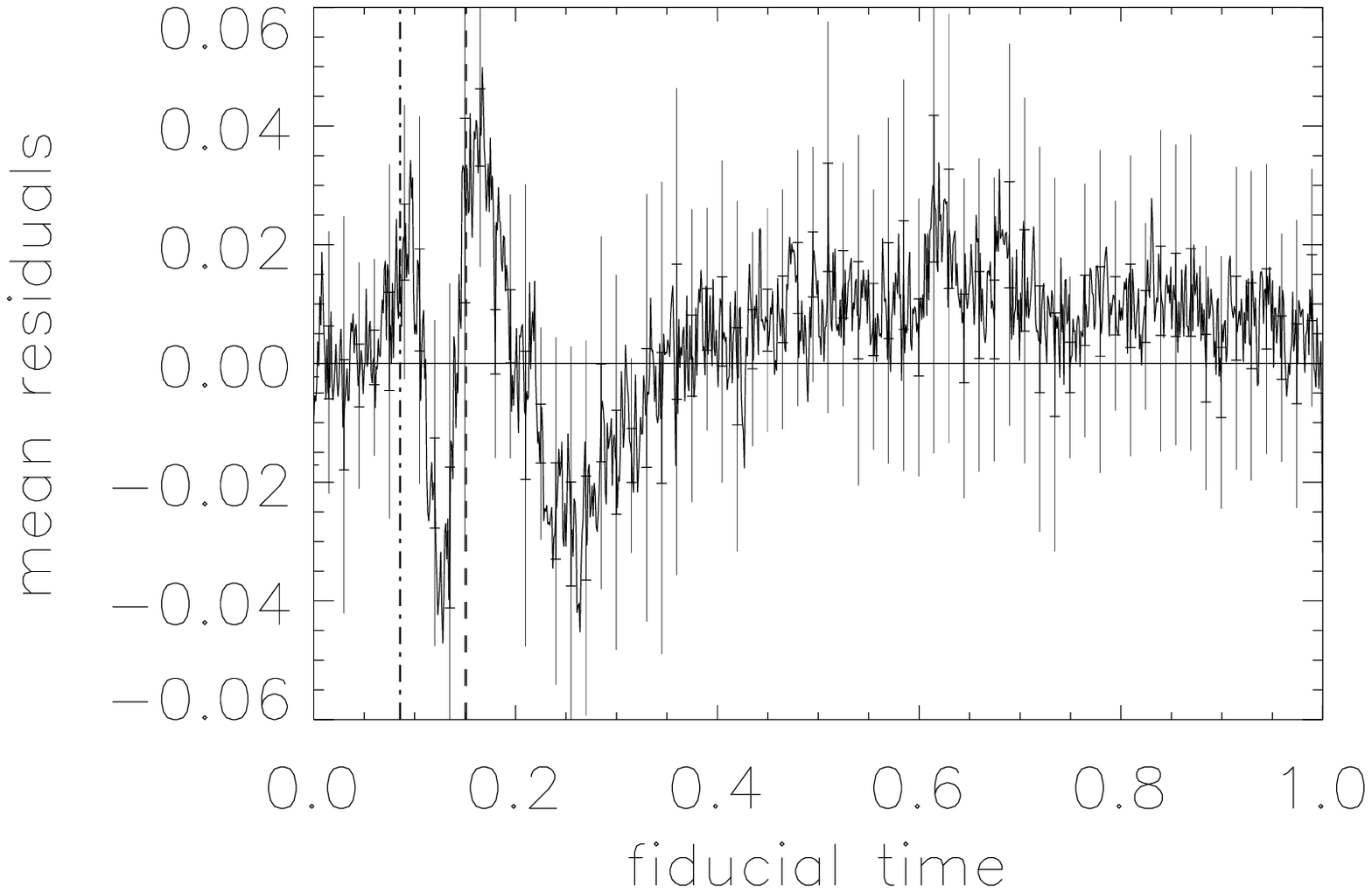}
  \caption{Pulse fit residuals for 13 very asymmetric (`v') pluses having $0.67 \le \kappa < 0.81$. The mean fitted pulse peak is found at fiducial time $\langle \Phi_{pk} \rangle=0.151$ (dashed line). \label{fig6}}
\end{figure}

\clearpage

\begin{figure}
  \includegraphics{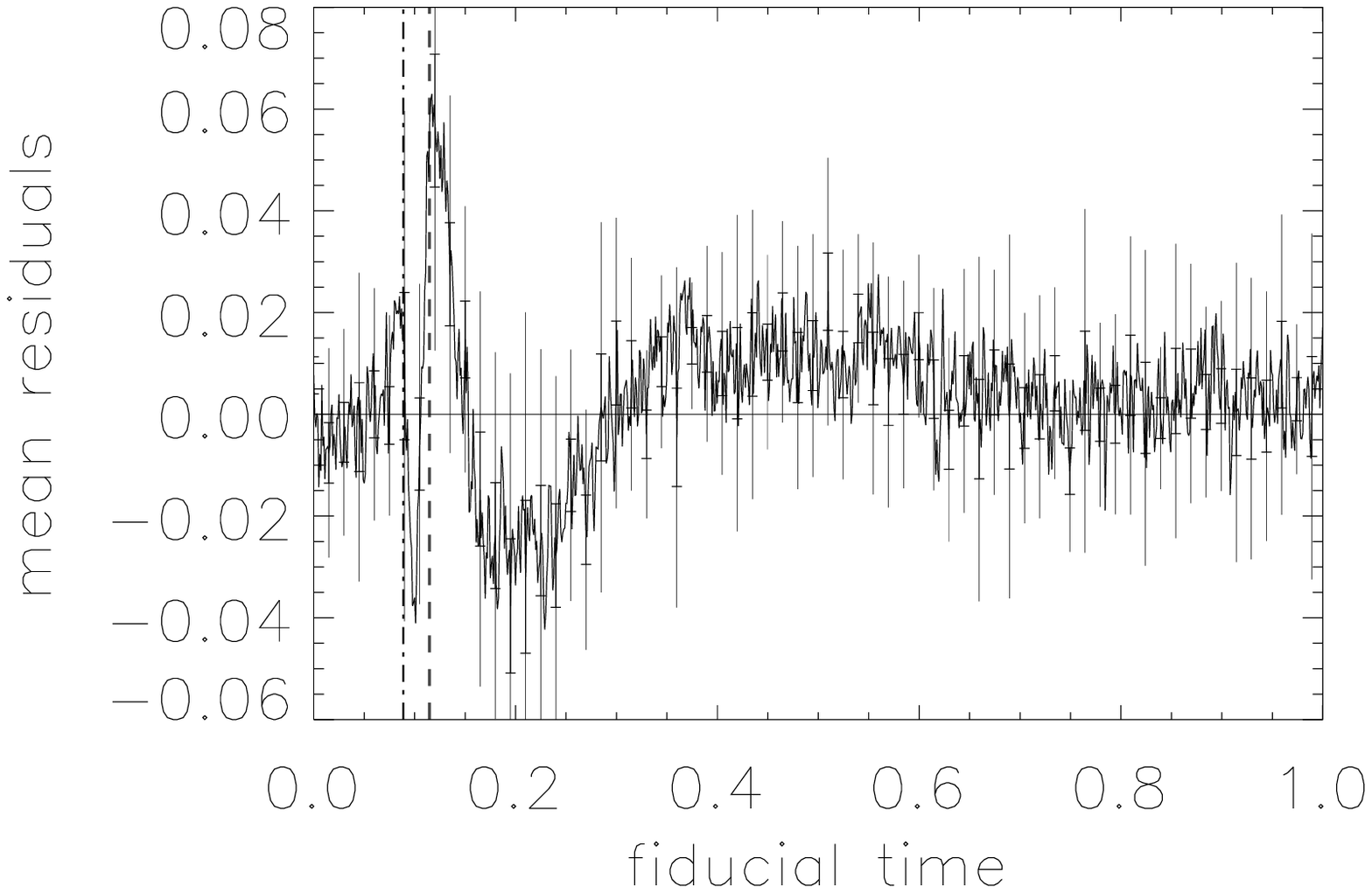}
  \caption{Pulse fit residuals for 13 extremely asymmetric (`x') pluses having  $\kappa \ge 0.81$. The mean fitted pulse peak is found at fiducial time $\langle \Phi_{pk} \rangle=0.115$ (dashed line). \label{fig7}}
\end{figure}

\clearpage

\begin{figure}
  \includegraphics{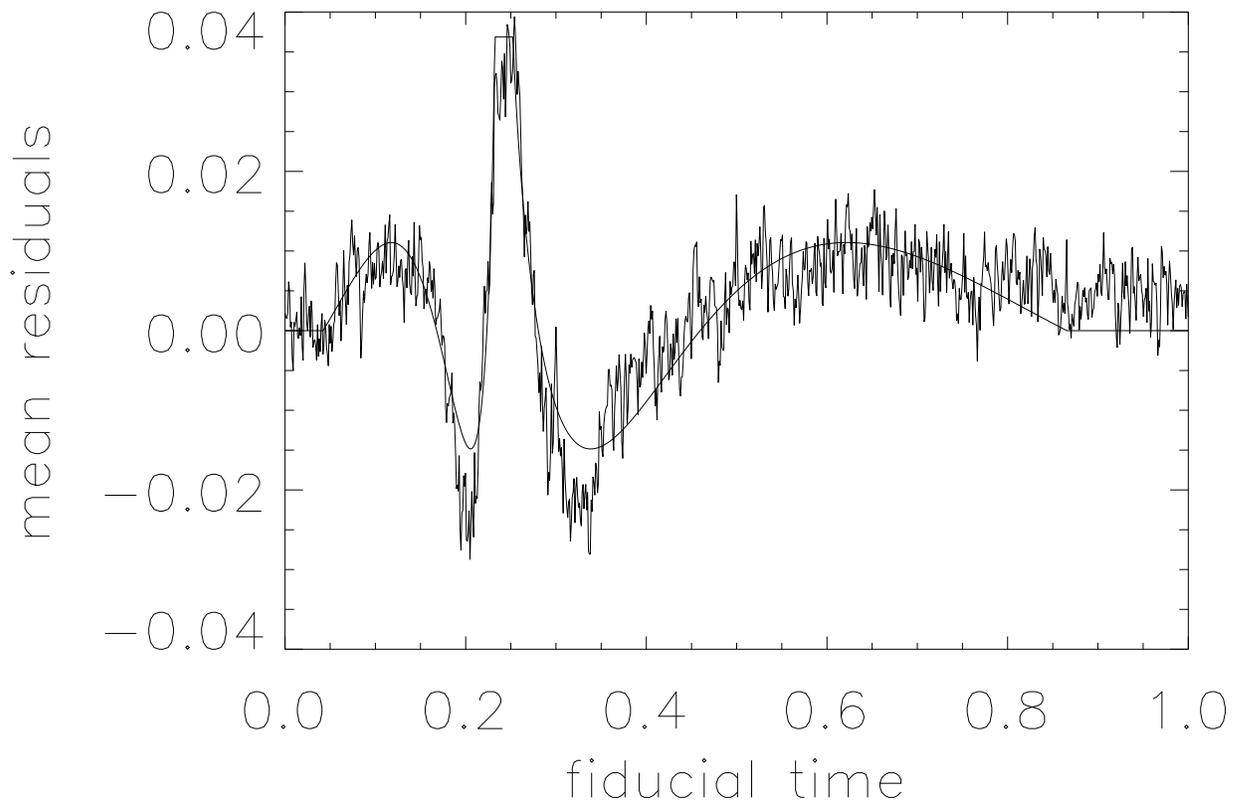}
  \caption{An empirical fit to the residual function shown in Figure 3. The functional form, fitting coefficients, and their uncertainties are described in the text.\label{fig8}}
\end{figure}

\clearpage

\begin{figure}
  \includegraphics{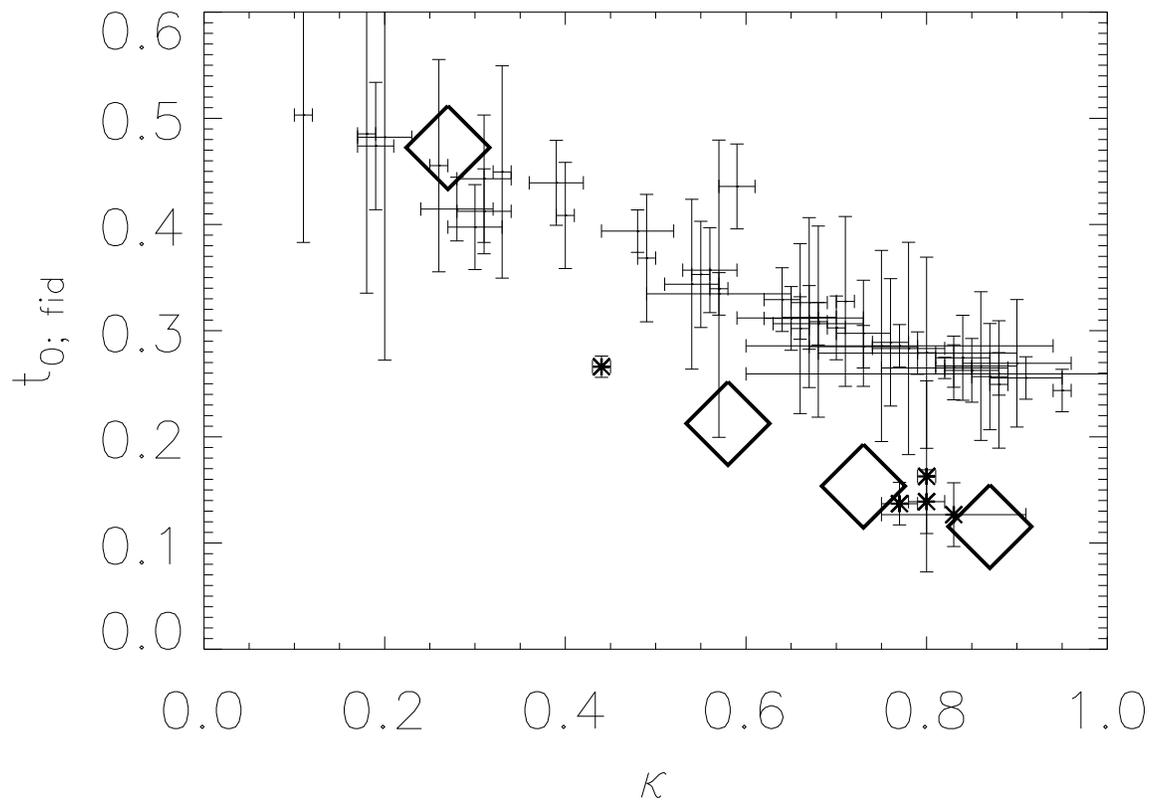}
  \caption{$t_{0; fid}$ vs. $\kappa$ for the \st{53} {\bf 50 calibration} pulses {\bf and 5 test pulses (asterisks) used} in this analysis. Template values for the subsets are identified by diamonds. \label{fig9}}
\end{figure}

\clearpage

\begin{figure}
  \includegraphics{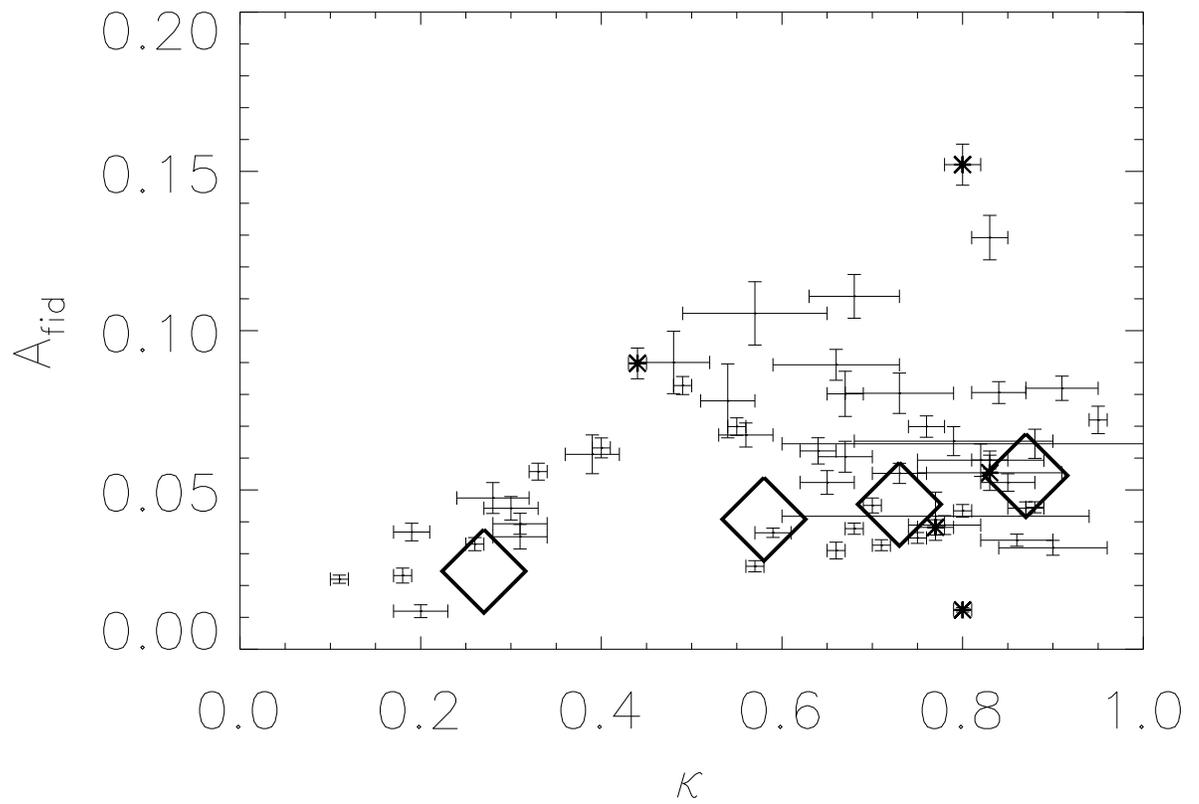}
  \caption{Residual amplitude $A$ vs. $\kappa$ for the 55 pulses in this analysis. Template values for the subsets are identified by diamonds.\label{fig10}}
\end{figure}

\clearpage

\begin{figure}
  \includegraphics{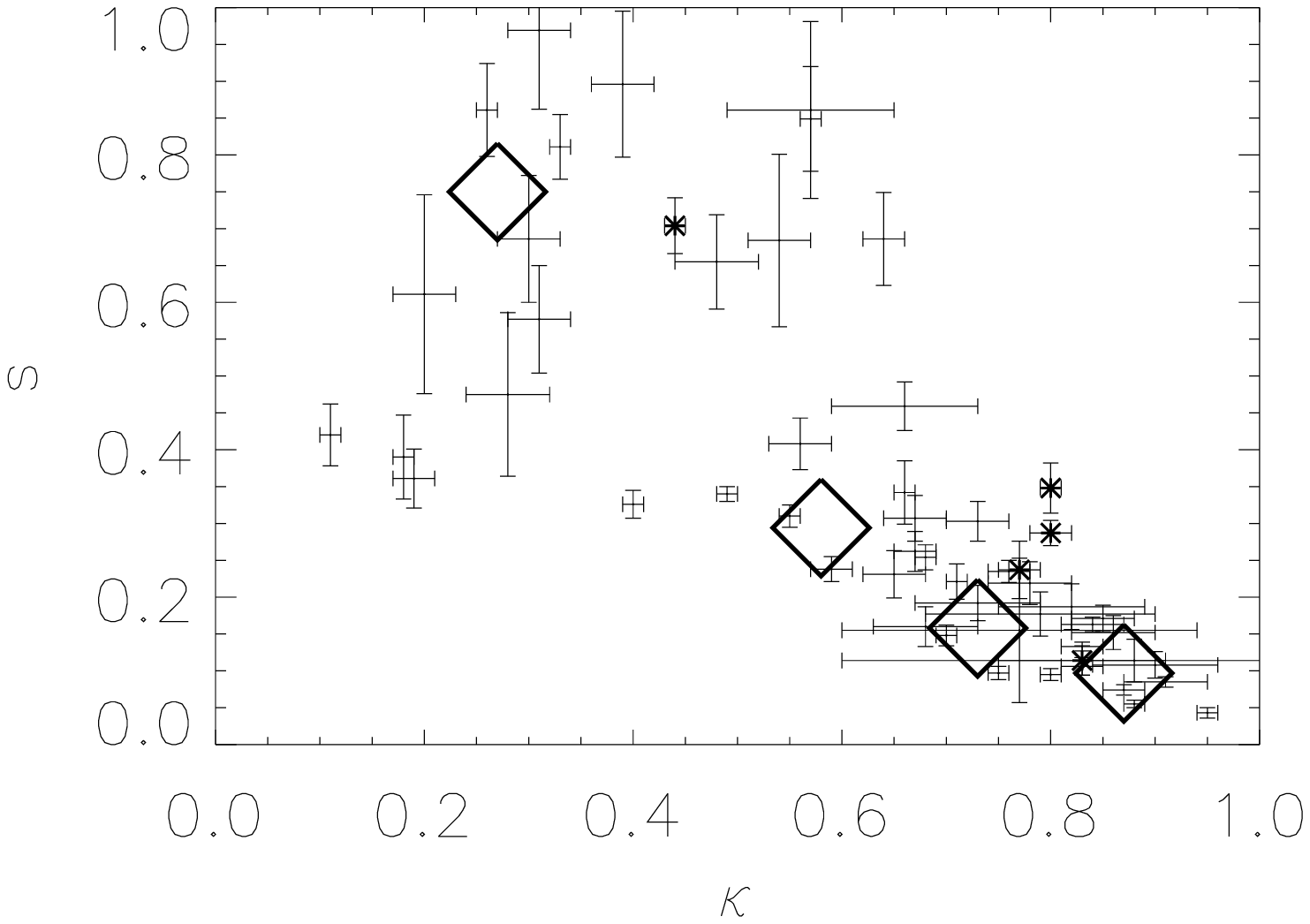}
  \caption{$\Omega_{\rm fid}$ vs. $\kappa$ for the 55 pulses in this analysis. Template values for the subsets are identified by diamonds.\label{fig11}}
\end{figure}

\clearpage

\begin{figure}
  \includegraphics{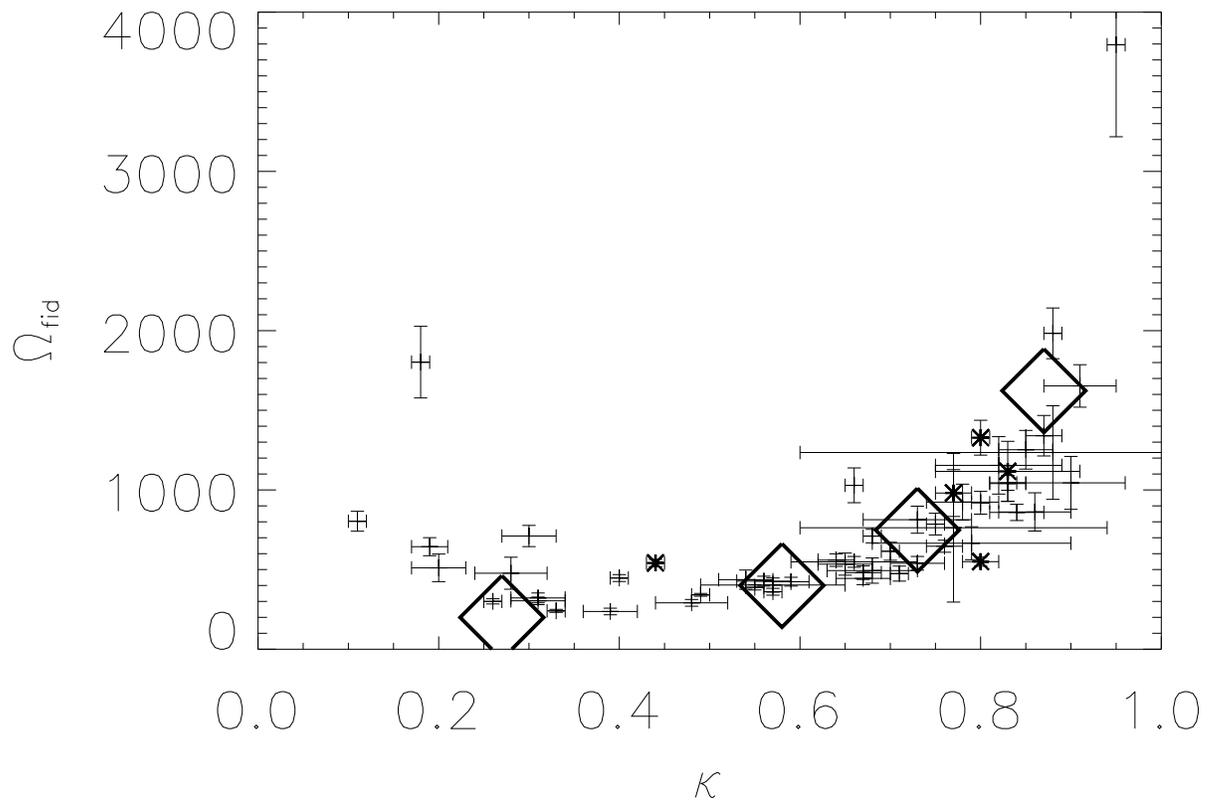}
  \caption{$s$ vs. $\kappa$ for the 55 pulses in this analysis. Template values for the subsets are identified by diamonds. \label{fig12}}
\end{figure}

\clearpage

\begin{figure}
\plottwo{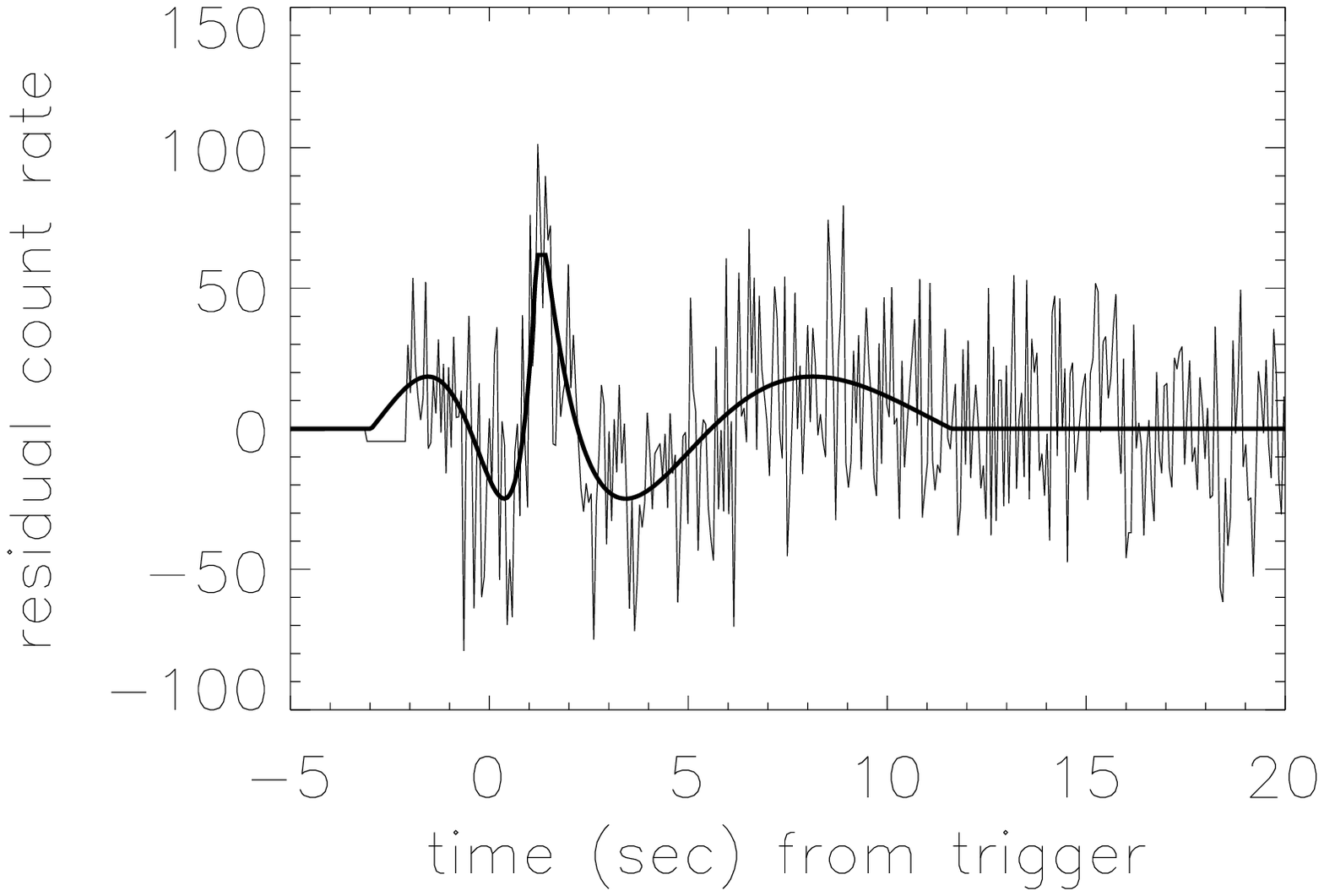}{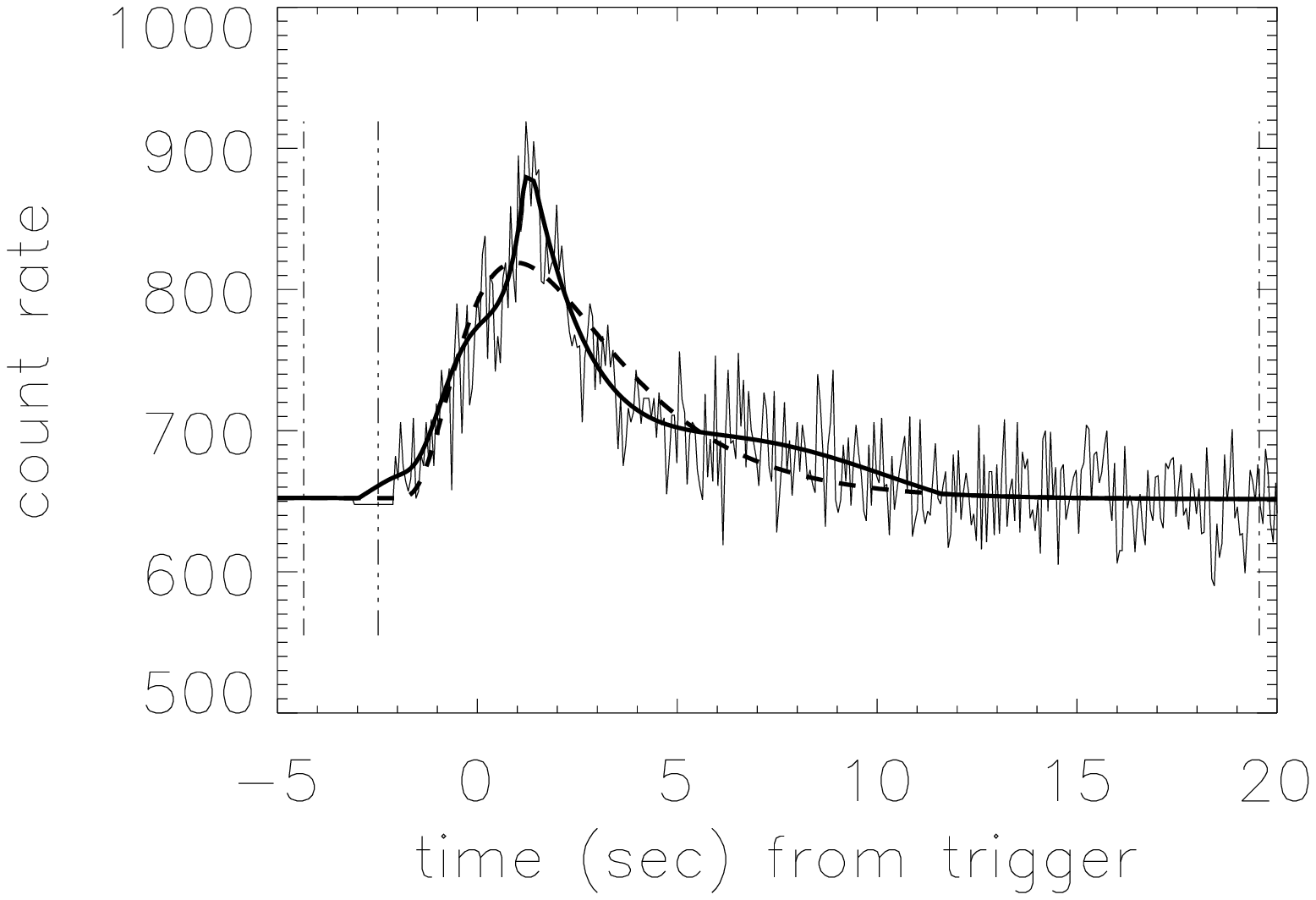}
\caption{The symmetric pulse in BATSE trigger 3026: (a) fit to the residuals, (b) fit to the residuals plus pulse model.  The first and last vertical lines on (b) mark the boundaries of the fiducial timescale, while the central vertical line marks the position of the formal start time of the \cite{nor05} pulse.
\label{fig13}}
\end{figure}

\clearpage

\begin{figure}
\plottwo{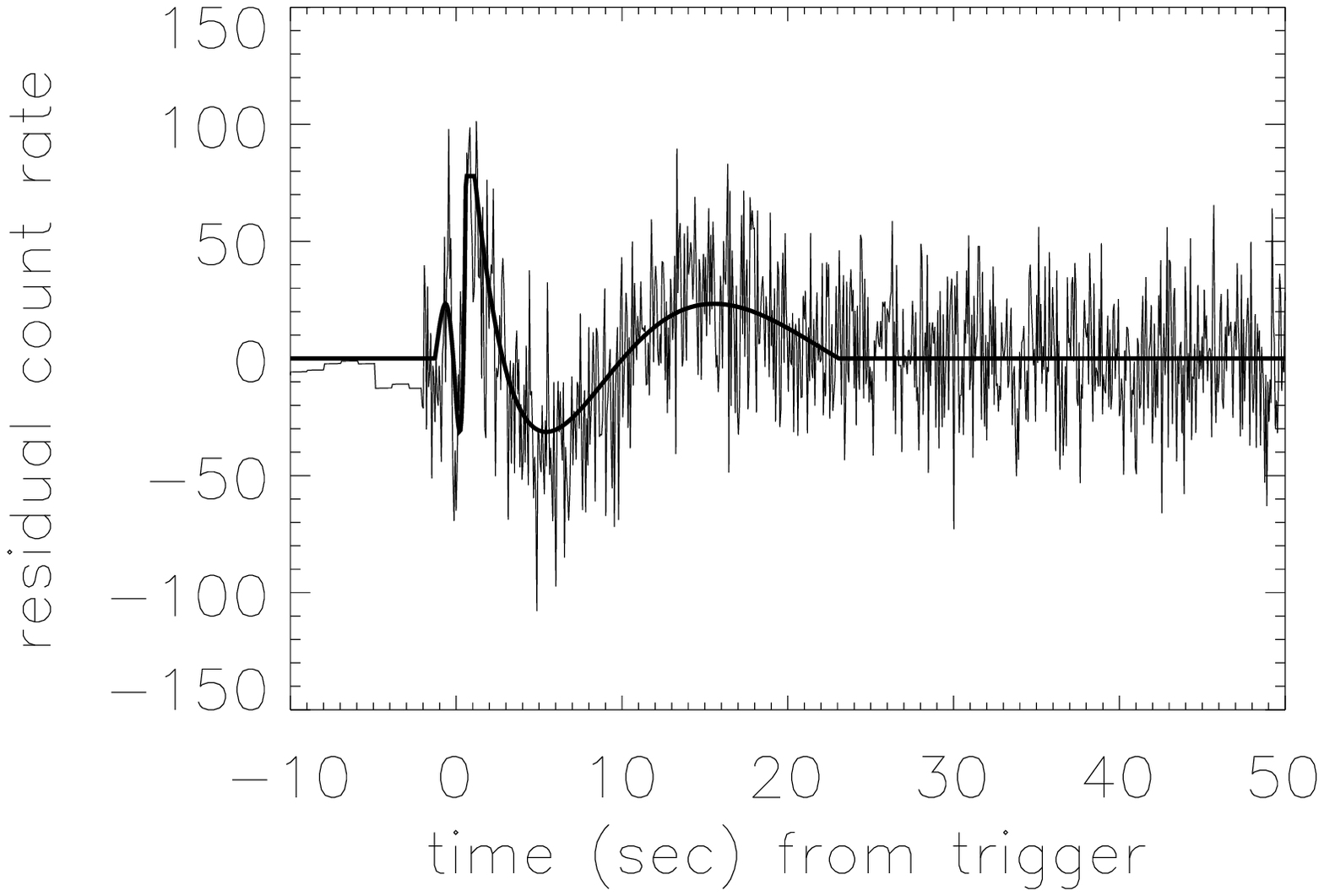}{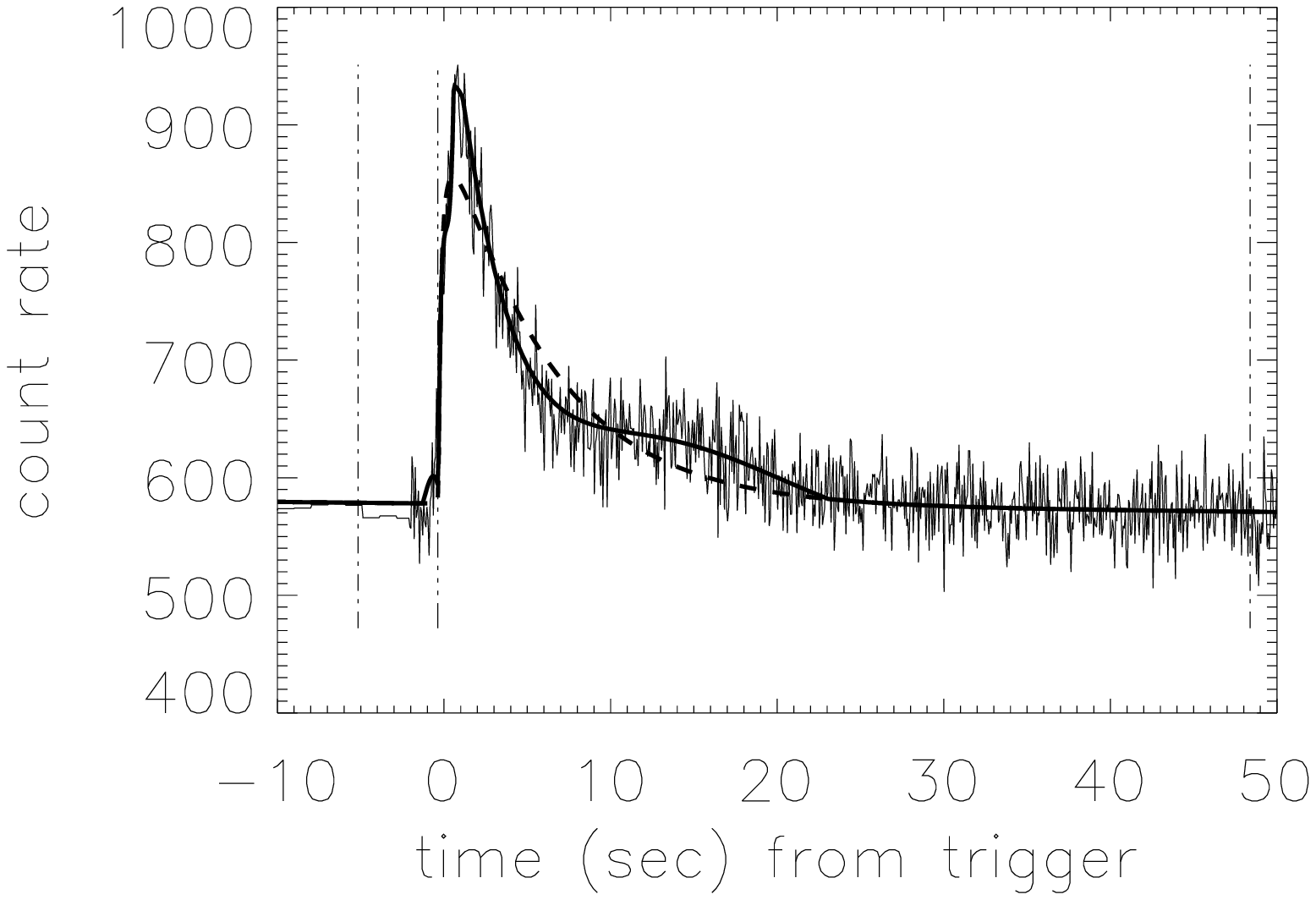}
\caption{The asymmetric pulse in BATSE trigger 3040: (a) fit to the residuals, (b) fit to the residuals plus pulse model. 
\label{fig14}}
\end{figure}

\clearpage

\begin{figure}
\plottwo{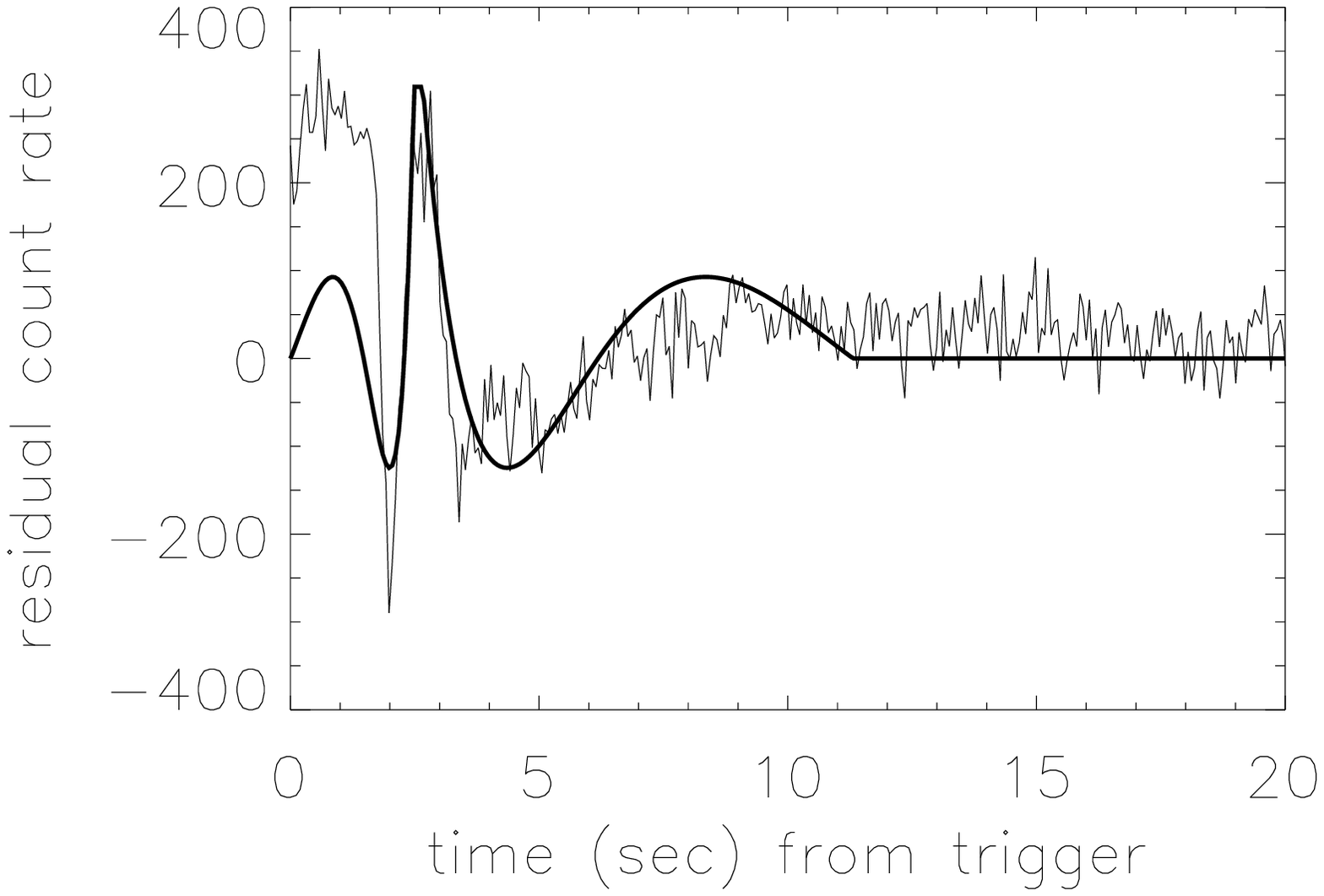}{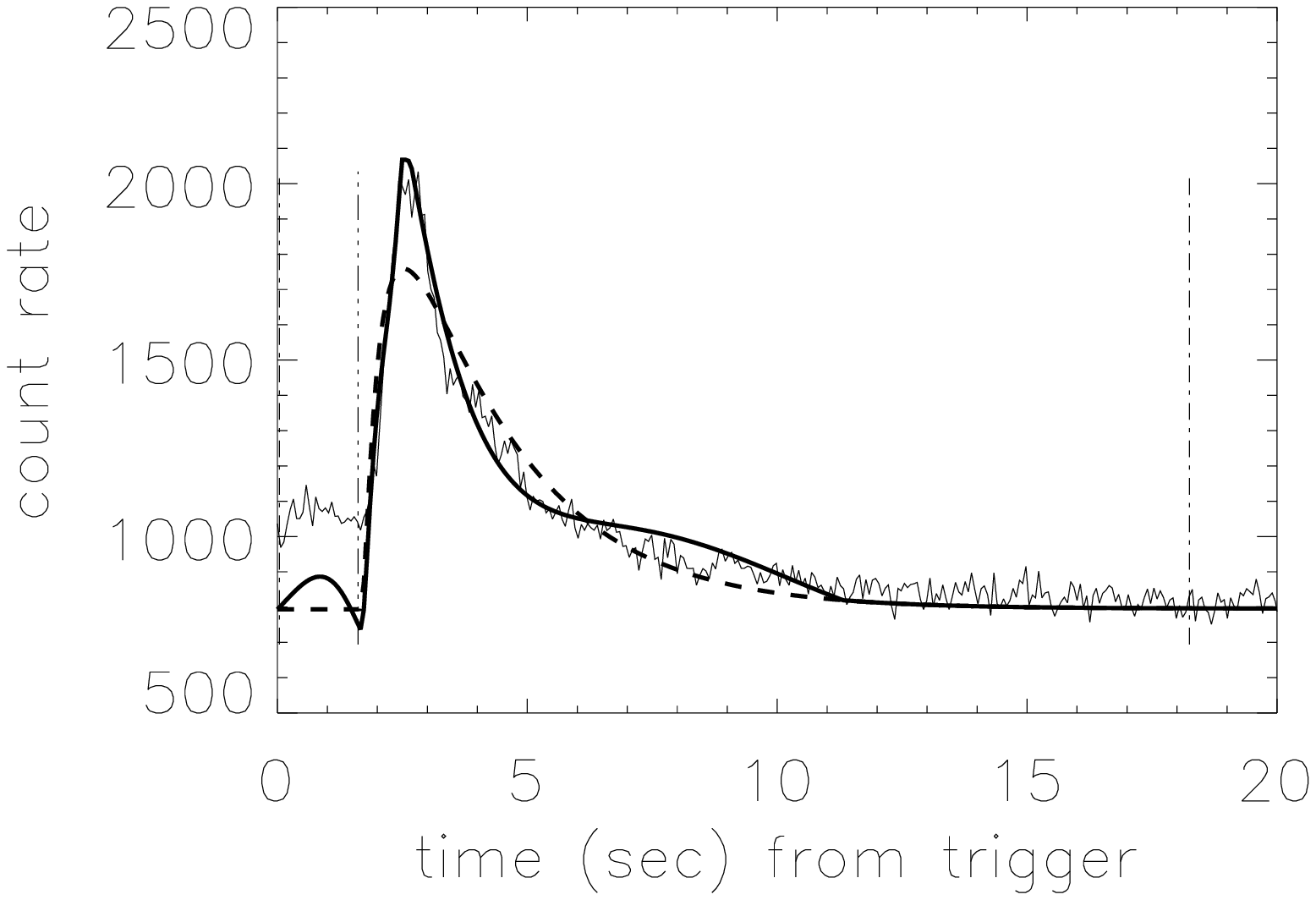}
\caption{The symmetric pulse in BATSE trigger 469: (a) fit to the residuals, (b) fit to the residuals plus pulse model. This pulse is a test pulse, and was not used in the calibration of the residual model. Note the very strong initial precursor. \label{fig15}}
\end{figure}

\clearpage

\begin{figure}
  \includegraphics{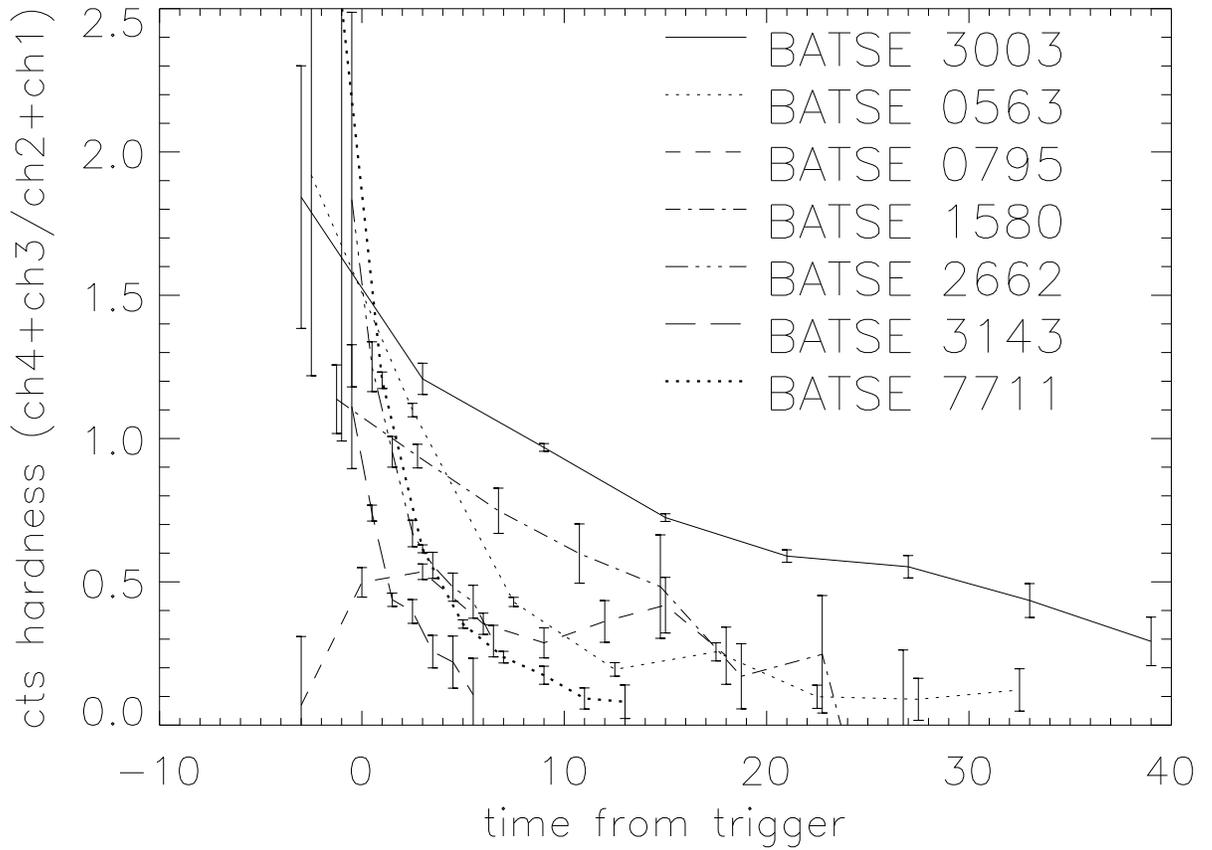}
\caption{Spectral evolution in seven typical GRB pulses. The preliminary fluctuation (prior to the trigger and generally prior to the pulse rise) almost always has the hardest spectral component. Additionally, pulses begin with different initial hardnesses, but all soften to similar values until times when the pulse intensity can no longer be detected.\label{fig16}}
\end{figure}

\clearpage

\begin{figure}
\plottwo{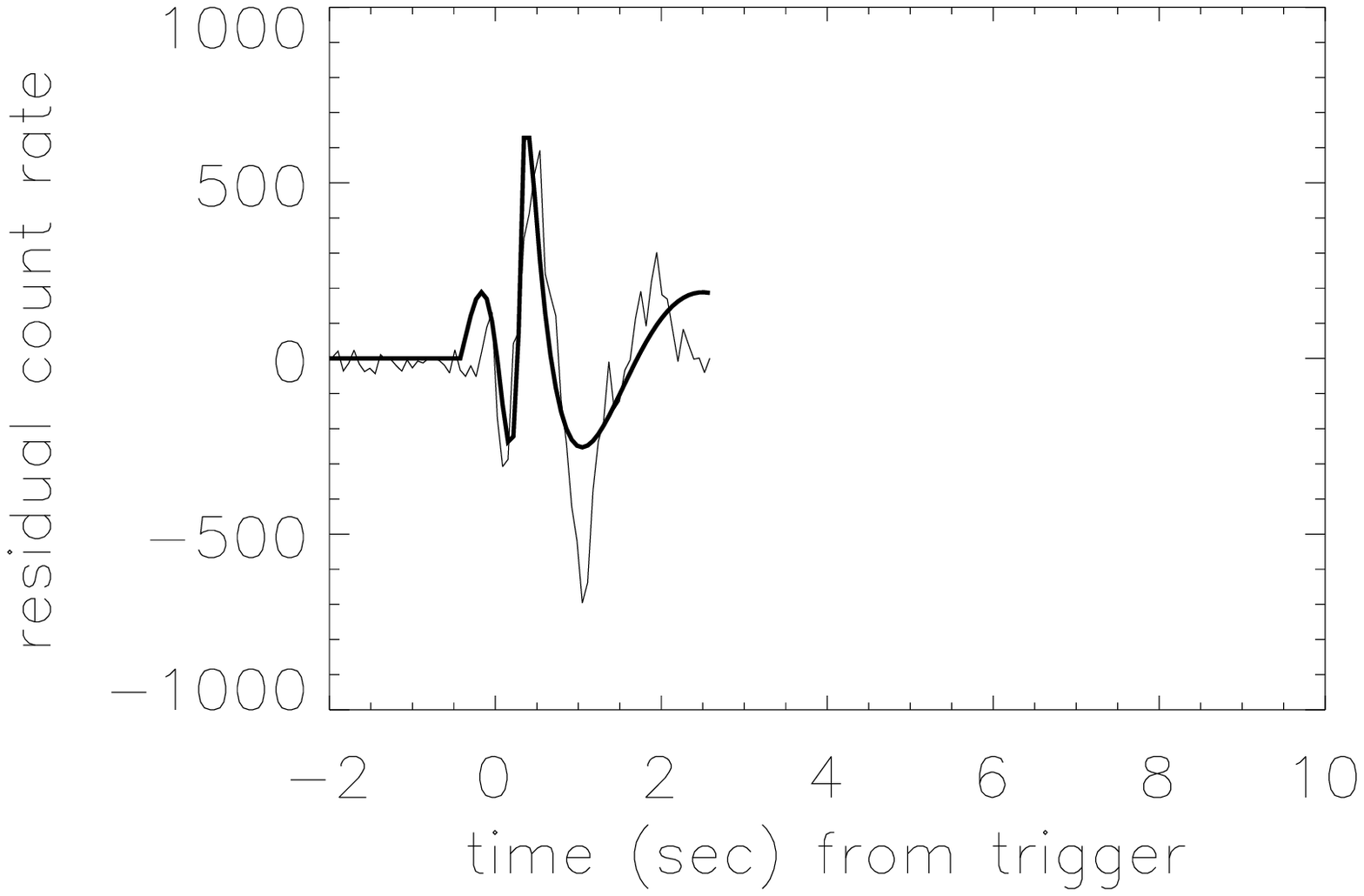}{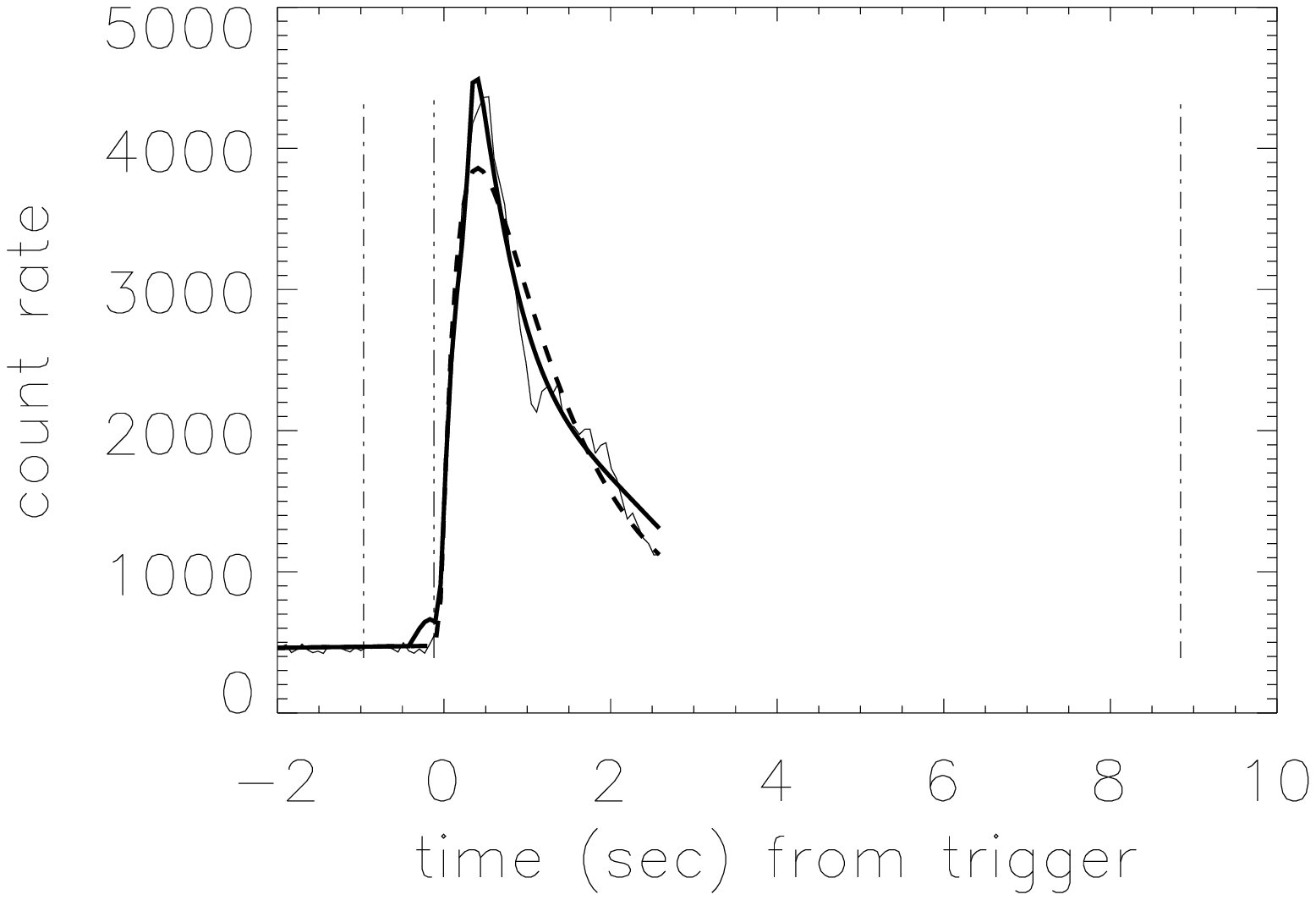}
\caption{The asymmetric trigger pulse in GRB 130427a: (a) fit to the residuals, (b) fit to the residuals plus pulse model. This long pulse is also used as a test pulse, rather than being used in the calibration of the residual model. \label{fig17}}
\end{figure}

\clearpage

\begin{figure}
\plottwo{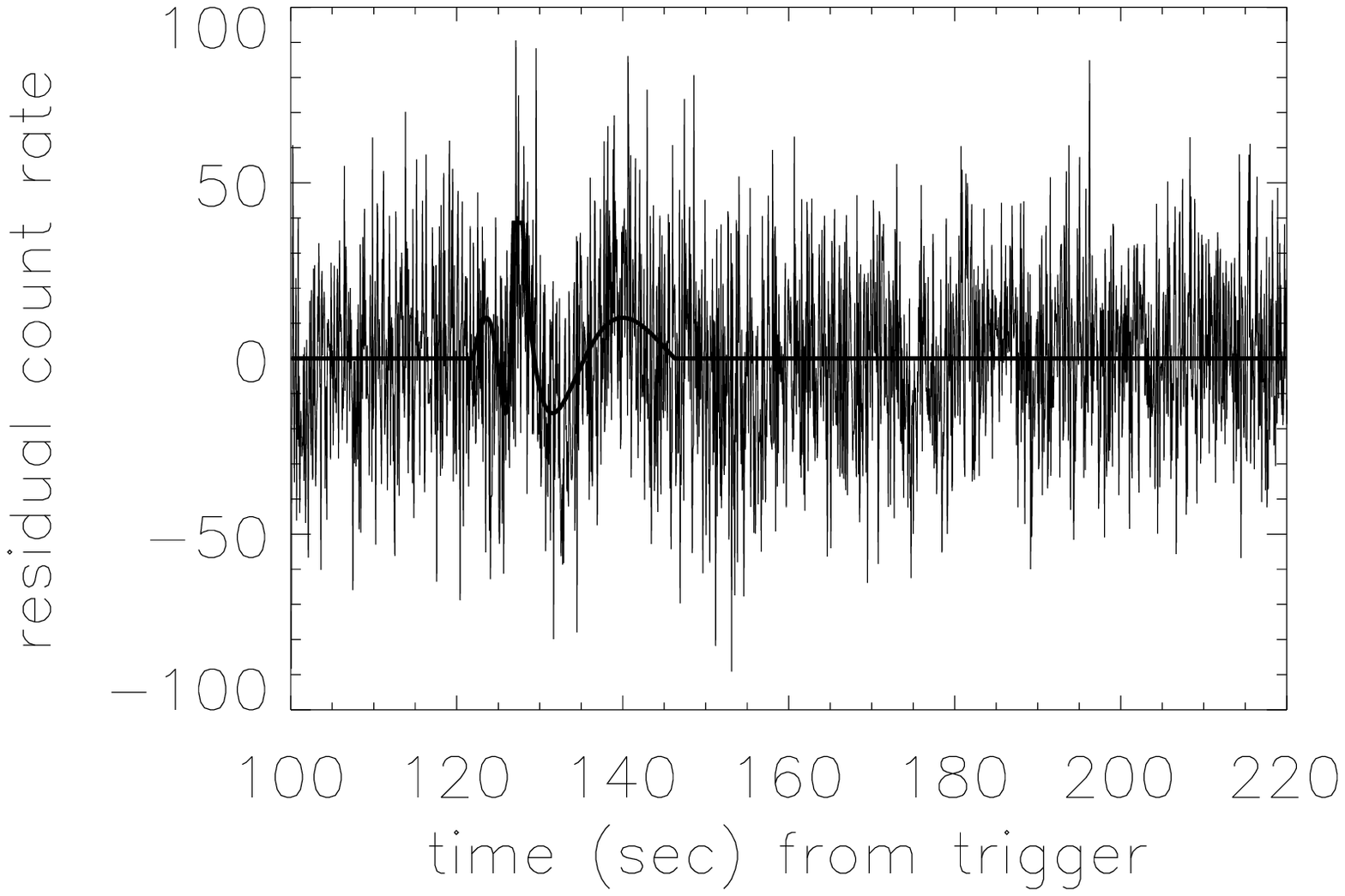}{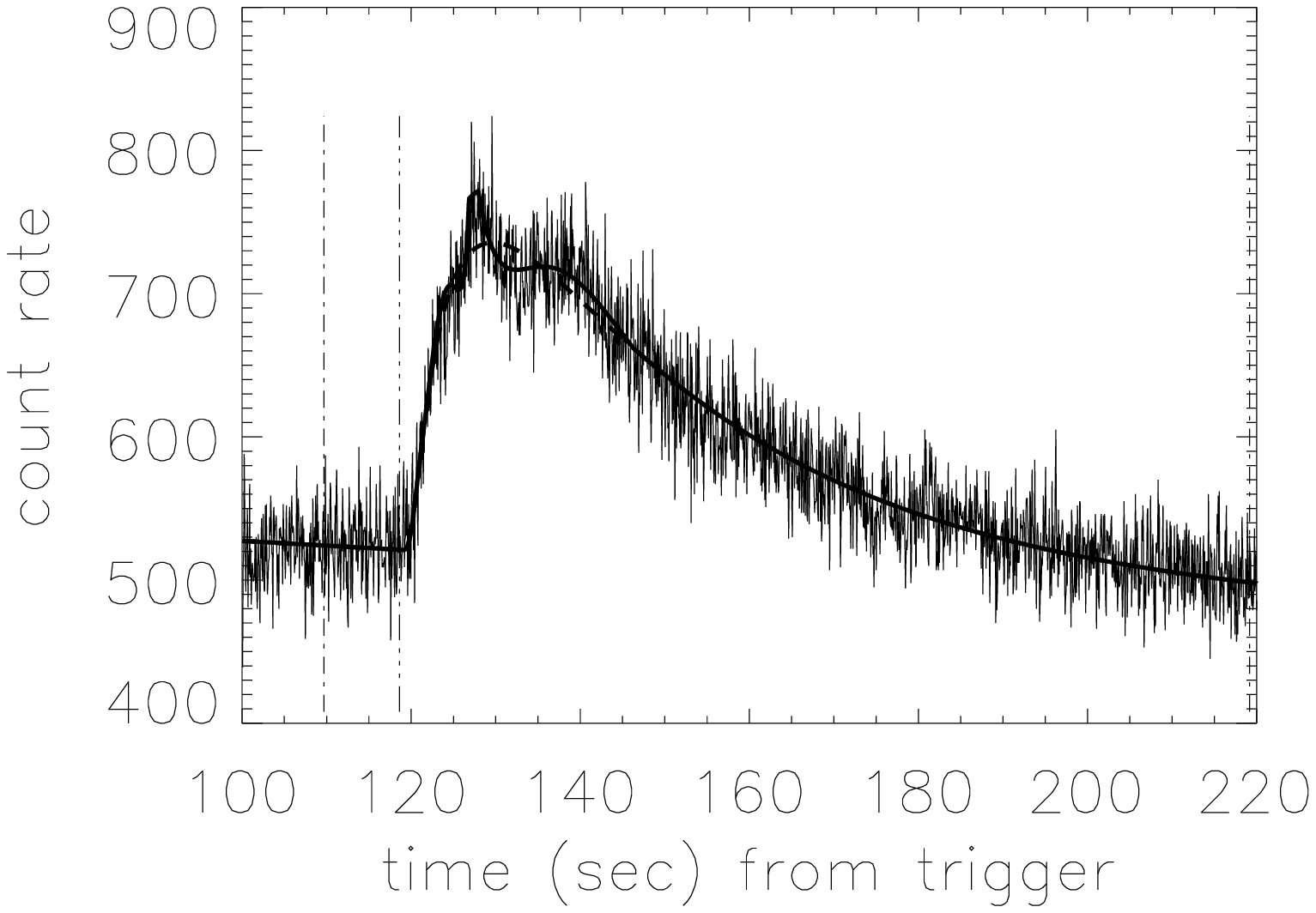}
\caption{The asymmetric final pulse in GRB 130427a: (a) fit to the residuals, (b) fit to the residuals plus pulse model. This long pulse is also used as a test pulse, rather than being used in the calibration of the residual model. \label{fig18}}
\end{figure}

\clearpage

\begin{figure}
\plottwo{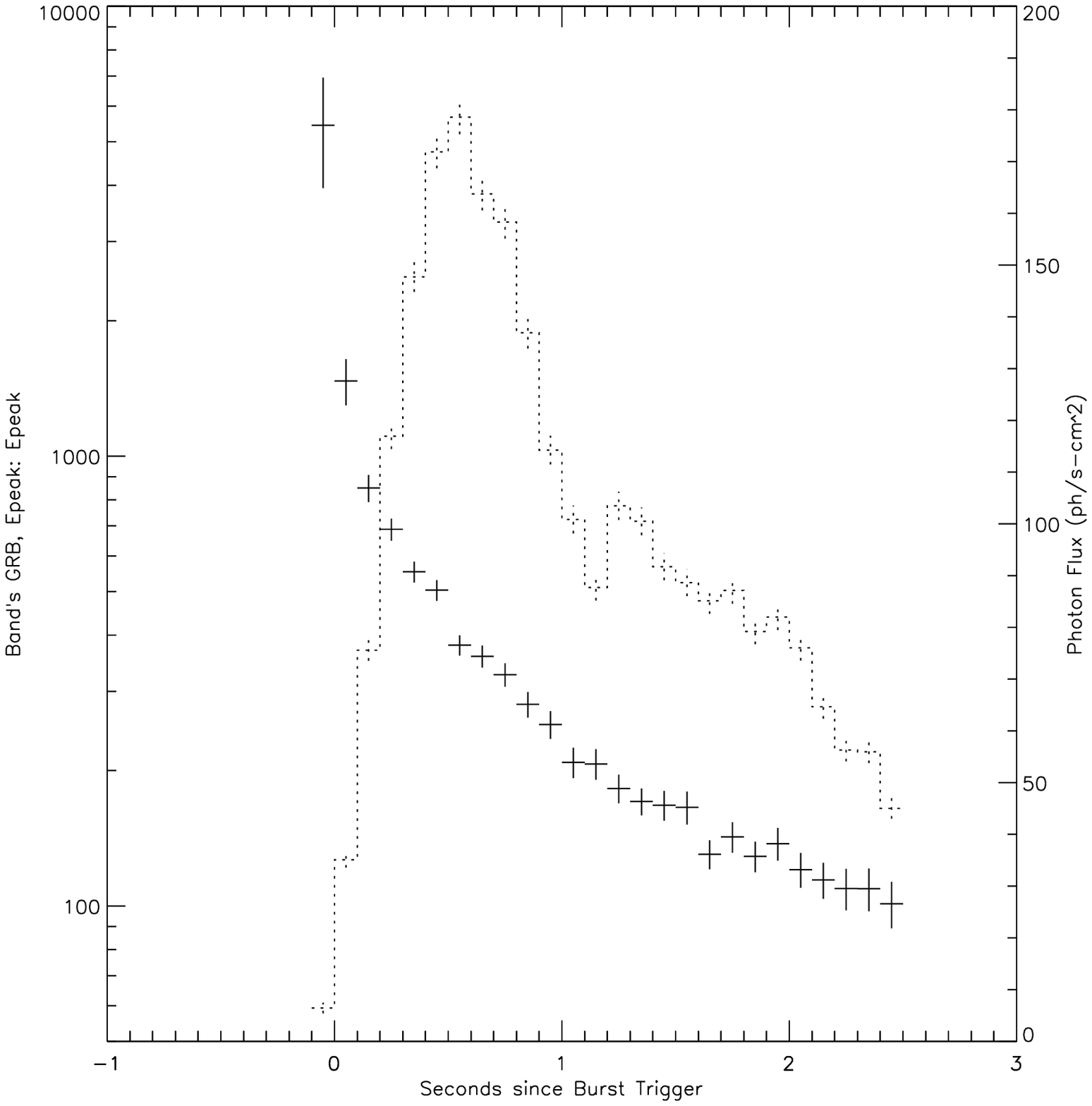}{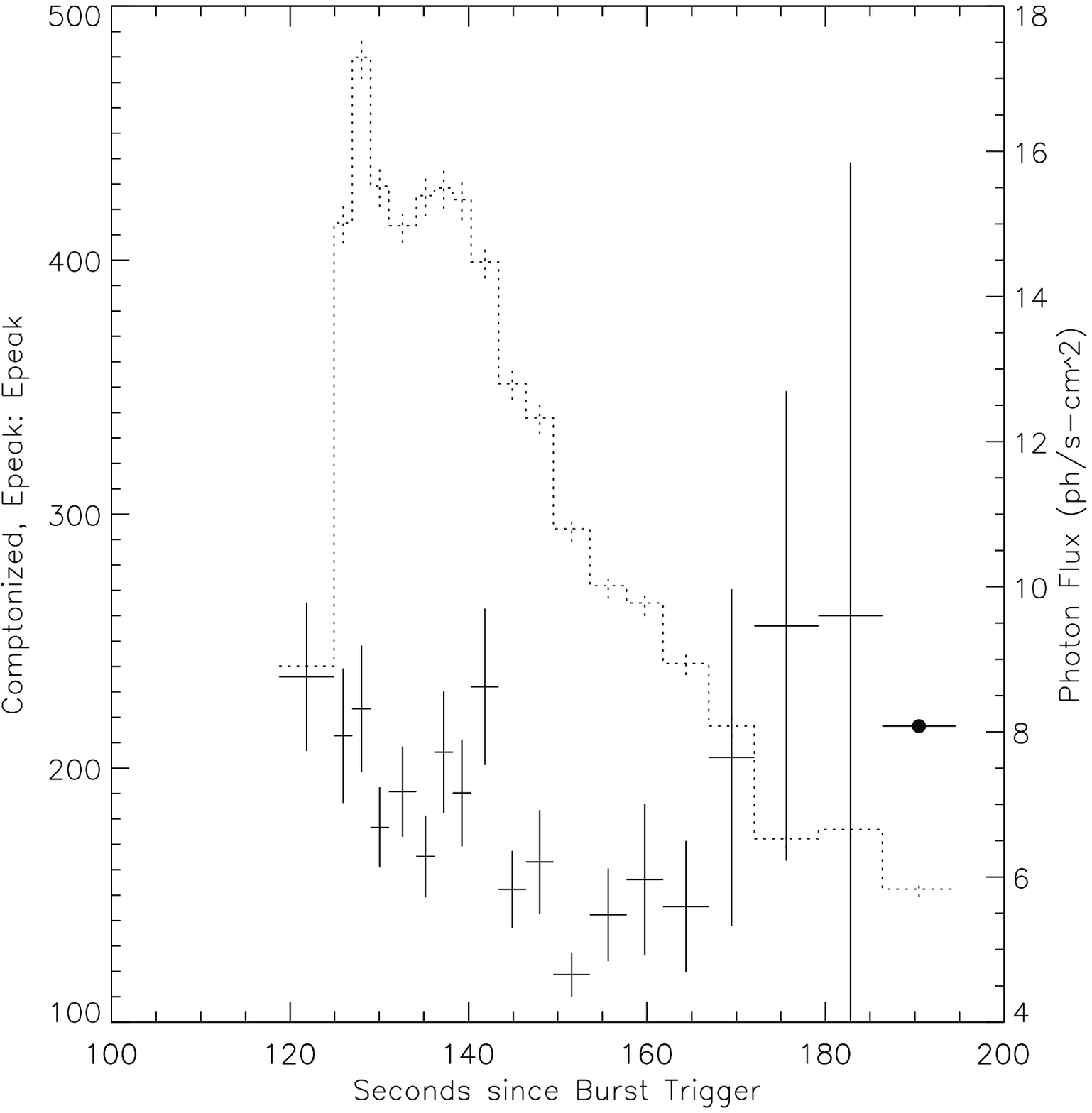}
\caption{Spectral evolution in the two fitted pulses in GRB 130427a: (a) the overlapped trigger pulse, (b) the long duration, low amplitude final pulse. The preliminary fluctuation (prior to the trigger and generally prior to the pulse rise) almost always has the hardest spectral component. Additionally, pulses begin with different initial hardnesses, but all soften to similar values until times when the pulse intensity can no longer be detected (the possible hardening at the end of the last pulse appears due to the onset of the afterglow phase).\label{fig19}}
\end{figure}

\clearpage

\begin{figure}
\includegraphics{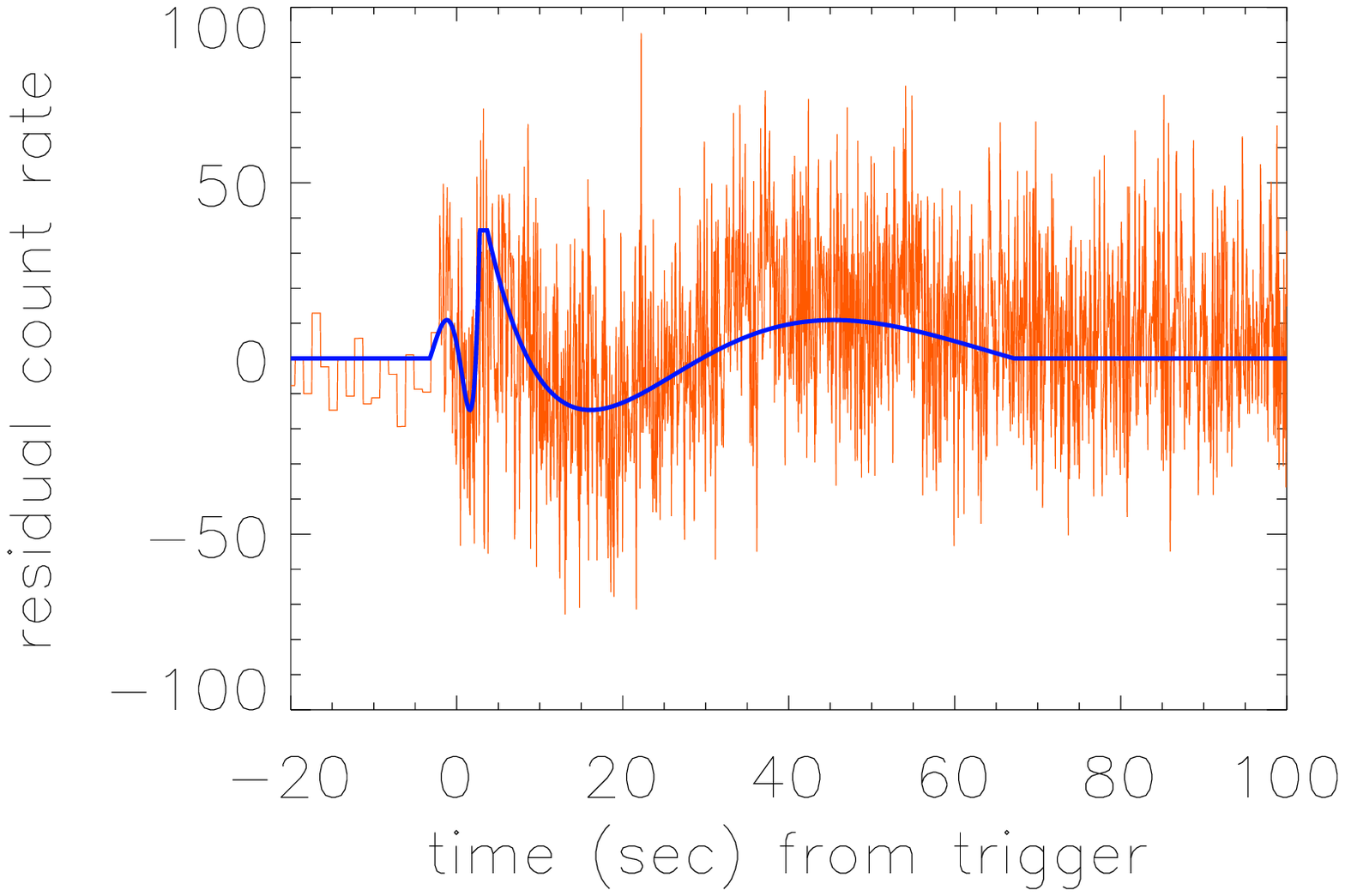}
\plottwo{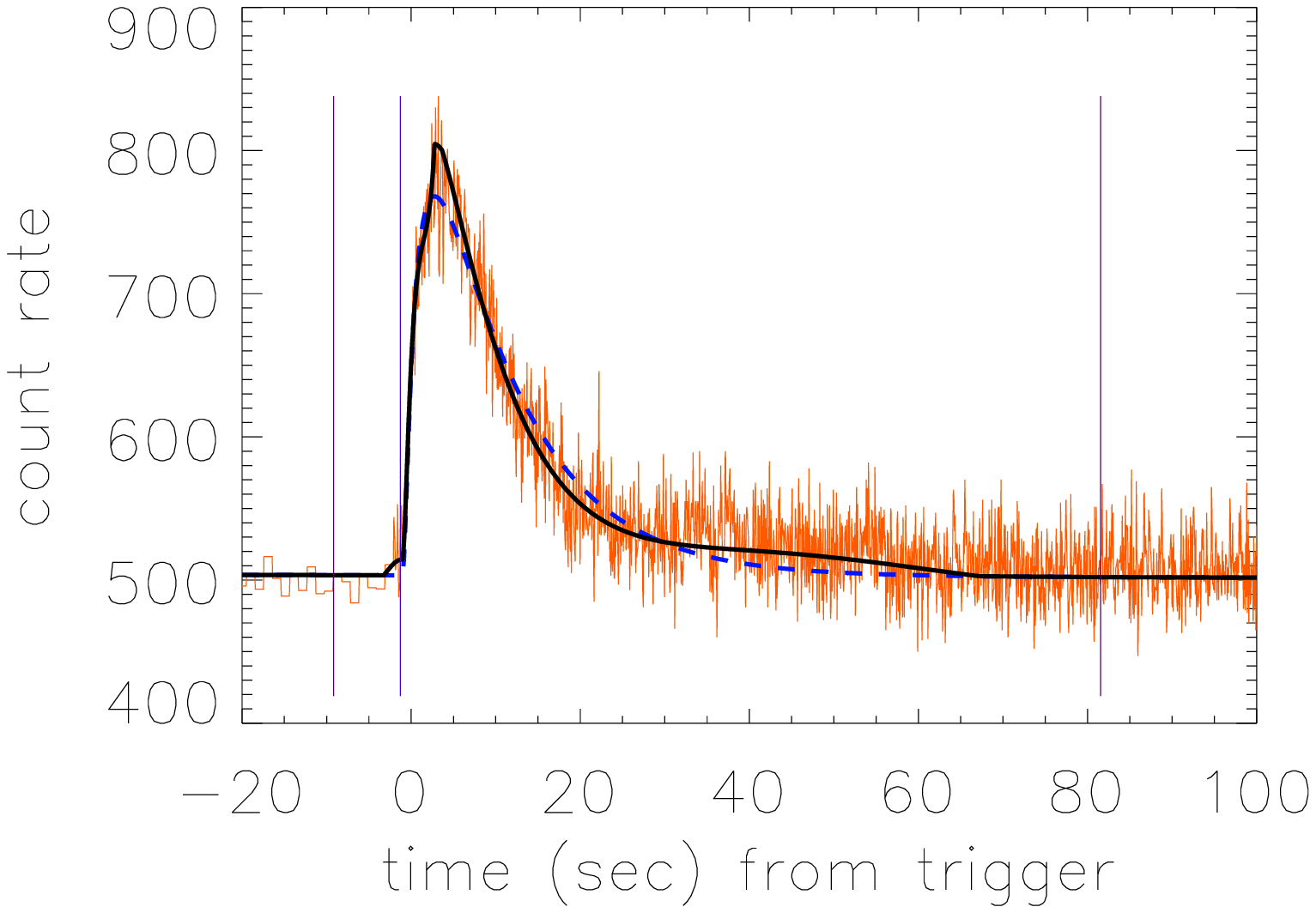}{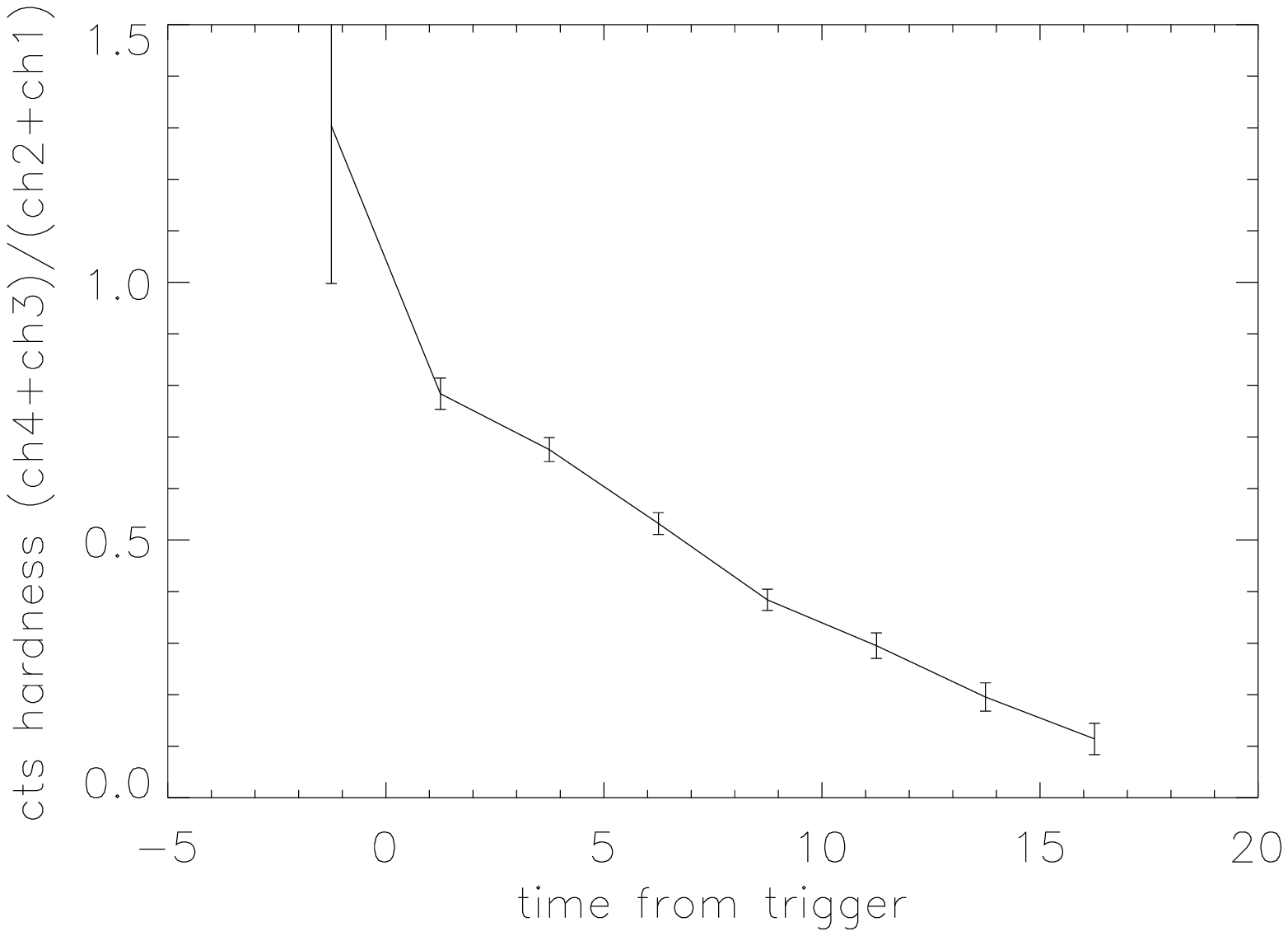}
\caption{{\em This is a sample of the figures to be included online as supplemental material.} BATSEÊ 332 (a) residual curve, (b) pulse fit, (c) hardness evolution.\label{fig20}}
\end{figure}

\clearpage

The following Figure numbers are for online figures. These may be currently found at

\texttt{http://authortools.aas.org/93249/FS20/figset.html}.

Figure 20. BATSEÊ 332 (a) residual curve, (b) pulse fit, (c) hardness evolution.

Figure 21. BATSEÊ 493 (a) residual curve, (b) pulse fit, (c) hardness evolution.

Figure 22. BATSEÊ 501 (a) residual curve, (b) pulse fit, (c) hardness evolution.

Figure 23. BATSEÊ 540 (a) residual curve, (b) pulse fit, (c) hardness evolution.

Figure 24. BATSEÊ 563 (a) residual curve, (b) pulse fit, (c) hardness evolution.

Figure 25. BATSEÊ 658 (a) residual curve, (b) pulse fit, (c) hardness evolution.

Figure 26. BATSE  673 (a) residual curve, (b) pulse fit, (c) hardness evolution.

Figure 27. BATSE  680 (a) residual curve, (b) pulse fit, (c) hardness evolution.

Figure 28. BATSE  711 (a) residual curve, (b) pulse fit, (c) hardness evolution.

Figure 29. BATSE  727 (a) residual curve, (b) pulse fit, (c) hardness evolution.

Figure 30. BATSEÊ 795 (a) residual curve, (b) pulse fit, (c) hardness evolution.

Figure 31. BATSEÊ 907 (a) residual curve, (b) pulse fit, (c) hardness evolution.

Figure 32. BATSEÊ 914 (a) residual curve, (b) pulse fit, (c) hardness evolution.

Figure 33. BATSE 1039 (a) residual curve, (b) pulse fit, (c) hardness evolution.

Figure 34. BATSE 1145 (a) residual curve, (b) pulse fit, (c) hardness evolution.

Figure 35. BATSE 1200 (a) residual curve, (b) pulse fit, (c) hardness evolution.

Figure 36. BATSE 1301 (a) residual curve, (b) pulse fit, (c) hardness evolution.

Figure 37. BATSE 1306 (a) residual curve, (b) pulse fit, (c) hardness evolution.

Figure 38. BATSE 1319 (a) residual curve, (b) pulse fit, (c) hardness evolution.

Figure 39. BATSE 1379 (a) residual curve, (b) pulse fit, (c) hardness evolution.

Figure 40. BATSE 1406 (a) residual curve, (b) pulse fit, (c) hardness evolution.

Figure 41. BATSE 1432 (a) residual curve, (b) pulse fit, (c) hardness evolution.

Figure 42. BATSE 1446 (a) residual curve, (b) pulse fit, (c) hardness evolution.

Figure 43. BATSE 1467 (a) residual curve, (b) pulse fit, (c) hardness evolution.

Figure 44. BATSE 1580 (a) residual curve, (b) pulse fit, (c) hardness evolution.

Figure 45. BATSE 1806 (a) residual curve, (b) pulse fit, (c) hardness evolution.

Figure 46. BATSE 1883 (a) residual curve, (b) pulse fit, (c) hardness evolution.

Figure 47. BATSE 2662 (a) residual curve, (b) pulse fit, (c) hardness evolution.

Figure 48. BATSE 2665 (a) residual curve, (b) pulse fit, (c) hardness evolution.

Figure 49. BATSE 2862 (a) residual curve, (b) pulse fit, (c) hardness evolution.

Figure 50. BATSE 2958p1 (a) residual curve, (b) pulse fit, (c) hardness evolution.

Figure 51. BATSE 3003 (a) residual curve, (b) pulse fit, (c) hardness evolution.

Figure 52. BATSE 3026 (a) residual curve, (b) pulse fit, (c) hardness evolution.

Figure 53. BATSE 3040 (a) residual curve, (b) pulse fit, (c) hardness evolution.

Figure 54. BATSE 3143 (a) residual curve, (b) pulse fit, (c) hardness evolution.

Figure 55. BATSE 3168 (a) residual curve, (b) pulse fit, (c) hardness evolution.

Figure 56. BATSE 3257 (a) residual curve, (b) pulse fit, (c) hardness evolution.

Figure 57. BATSE 7567 (a) residual curve, (b) pulse fit, (c) hardness evolution.

Figure 58. BATSE 7614 (a) residual curve, (b) pulse fit, (c) hardness evolution.

Figure 59. BATSE 7638 (a) residual curve, (b) pulse fit, (c) hardness evolution.

Figure 60. BATSE 7711 (a) residual curve, (b) pulse fit, (c) hardness evolution.

Figure 61. BATSE 7775 (a) residual curve, (b) pulse fit, (c) hardness evolution.

Figure 62. BATSE 7843 (a) residual curve, (b) pulse fit, (c) hardness evolution.

Figure 63. BATSE 7903 (a) residual curve, (b) pulse fit, (c) hardness evolution.

Figure 64. BATSE 7989 (a) residual curve, (b) pulse fit, (c) hardness evolution.

Figure 65. BATSE 8112 (a) residual curve, (b) pulse fit, (c) hardness evolution.

Figure 66. BATSE 8121 (a) residual curve, (b) pulse fit, (c) hardness evolution.

Figure 67. GRB081224 (a) residual curve, (b) pulse fit, (c) hardness evolution.

Figure 68. GRB100707a (a) residual curve, (b) pulse fit, (c) hardness evolution.

Figure 69. GRB120326a (a) residual curve, (b) pulse fit, (c) hardness evolution.

Figure 70. BATSE 0469 (a) residual curve, (b) pulse fit, (c) hardness evolution.

Figure 71. BATSE 3164 (a) residual curve, (b) pulse fit, (c) hardness evolution.

Figure 72. BATSE 2958p2 (a) residual curve, (b) pulse fit, (c) hardness evolution.

\end{document}